\documentclass[aps,english,prl,floatfix,amsmath,superscriptaddress,tightenlines,twocolumn]{revtex4-2}
\usepackage{mathtools, amssymb}
\usepackage{graphicx}
\usepackage{tikz}
\usepackage{tikz-3dplot}
\usetikzlibrary{spy}
\usetikzlibrary{arrows.meta}
\usetikzlibrary{calc}
\usetikzlibrary{decorations.pathreplacing,calligraphy}
\usepackage{intcalc}
\usepackage[caption=false]{subfig}
\usepackage[shortlabels]{enumitem}
\usepackage{soul}
\usepackage[utf8]{inputenc}
\usepackage{xcolor}
\usepackage{amsthm}
\usepackage{float}
\usepackage{comment}
\usepackage{braket}
\usepackage{ulem}

\newcommand{\tr}{\textrm{Tr}}
\newcommand{\bo}{\mathbb{I}}

\newcommand{\mN}{\mathcal{N}}

\begin{document}

\title{Approximate quantum error correcting codes from conformal field theory}

\author{Shengqi Sang}
\affiliation{Perimeter Institute for Theoretical Physics, Waterloo, Ontario N2L 2Y5, Canada}
\affiliation{University of Waterloo, Waterloo, Ontario, N2L 3G1, Canada}

\author{Timothy H. Hsieh}
\affiliation{Perimeter Institute for Theoretical Physics, Waterloo, Ontario N2L 2Y5, Canada}

\author{Yijian Zou}
\email{yzou@perimeterinstitute.ca}
\affiliation{Perimeter Institute for Theoretical Physics, Waterloo, Ontario N2L 2Y5, Canada}

\begin{abstract}

The low-energy subspace of a conformal field theory (CFT) can serve as a quantum error correcting code, with important consequences in holography and quantum gravity.   We consider generic 1+1D CFT codes under extensive local dephasing channels and analyze their error correctability in the thermodynamic limit. We show that (i) there is a finite decoding threshold if and only if the minimal nonzero scaling dimension in the fusion algebra generated by the jump operator of the channel is larger than $1/2$ and (ii) the number of protected logical qubits $k \geq \Omega( \log \log n)$, where $n$ is the number of physical qubits.  As an application, we show that the one-dimensional quantum critical Ising model has a finite threshold for certain types of dephasing noise.  Our general results also imply that a CFT code with continuous symmetry saturates a bound on the recovery fidelity for covariant codes.


\end{abstract}

\maketitle

\textit{Introduction.--} Quantum information is fragile and can be lost when subject to decoherence. In order to robustly store and manipulate quantum information against noise, one embeds the logical qubits into a larger set of physical qubits, forming quantum error correcting codes (QECC) \cite{Shor1995,gottesman1997stabilizer}. QECCs provide promising routes to fault-tolerant quantum computing \cite{aharonov1999faulttolerant,Knill_1998} and have been recently realized on quantum simulators \cite{Google2023}. On the theoretical side, code properties of QECCs provide insight into patterns of long-range quantum many-body entanglement, connecting quantum information with condensed matter physics \cite{Dennis_2002,Kitaev2003} and quantum gravity \cite{Almheiri_2015,Pastawski_2015,Harlow_2017}. More recently, studies have suggested a close relationship between mixed-state topological phases and the ability of the QECCs to correct certain errors \cite{fan2024diagnostics,sang2023mixedstate}. It is thus important, both from a practical and a theoretical point of view, to understand which physical systems can serve as QECCs and which errors are correctable given the system or code.   

Most previous studies focus on stabilizer codes \cite{gottesman1997stabilizer}. 
The stabilizer formalism offers a natural way for decoding, where in some cases the optimal decoder can be found \cite{Dennis_2002}. However, the notion of QECC goes far beyond stabilizer codes. In order to define a QECC, one may specify a $D$-dimensional code subspace spanned by a set of mutually orthogonal codeword states, $\{|\phi_{\alpha}\rangle,~\alpha = 1,2, \cdots, D\}$, where $D=2^k$ to encode $k$ logical qubits. In the seminal work \cite{Brandao_2019}, the codeword states are chosen as the eigenstates of a local Hamiltonian. The code properties can then be related to physical properties of the system in an energy window. Another example that goes beyond stabilizers is AdS/CFT correspondence, where the code subspace is the CFT low-energy subspace corresponding to bulk graviton excitations. Such codes have recently been constructed explicitly \cite{Steinberg_2023,steinberg2024farperfectquantumerror}, where the code properties reveal important aspects of quantum gravity. Going beyond stabilizer codes poses significant challenges to determine error correctability as there are usually no explicit decoders. We note that there are existing information-theoretic criteria for the approximate correctability of errors \cite{yi2023complexity,Beny_2010,zheng2024nearoptimal} that have been used in a variety of contexts.


In this work, we show that {\it generic} CFTs furnish approximate quantum error correcting codes \cite{Leung1997,crepeau2005approximate}, where the code subspace is given by the low-energy subspace of the CFT. We note that for a fine-grained class of CFTs, certain exact codes can be derived using a different construction from this work \cite{Dymarsky_2021}. As opposed to the holographic correspondence, here the CFT can be realized by a simple spin chain at criticality, such as the transverse field Ising model. We analyze the decodability of CFT codes under a finite-time evolution of a translation-invariant Lindbladian, such as uniform dephasing noise. Our main result is that errors can be corrected at a finite threshold if and only if $\Delta_{\min}>1/2$, where $\Delta_{\min}$ is the minimal nonzero scaling dimension of CFT operators in fusion algebra generated by the Lindbladian jump operator. A natural corollary is that a CFT code can correct uniform depolarization noise at a finite threshold if and only if the algebra generated by the jump operator only contains CFT operators with scaling dimension larger than $1/2$. In addition, we show that for any CFT code the number of protected logical qubits $k\geq \Omega(\log \log n)$, where $n$ is the number of physical qubits.
As an example, we analyze the Ising CFT code realized by the low-energy subspace of the critical transverse-field Ising model (TFIM) in one dimension. We show that for dephasing noise, the code does not have a finite threshold for $X$ dephasing but can correct $Y$ and $Z$ dephasing up to maximum strength. 

Our work is in part inspired by recent developments in decoherence-induced mixed-state phases 
\cite{fan2023diagnostics, bao2023mixed, lee2023quantum, zou2023channeling, de2022symmetry, ma2023average, zhang2022strange, ma2023topological, lu2023mixed, chen2023symmetry, chen2023separability, lessa2024mixed, chen2024unconventional, lee2022symmetry, sala2024quantum,lee2024exact, sohal2024noisy, ellison2024towards, wang2023intrinsic, lessa2024strongtoweak, sala2024spontaneous, sang2024stability}.
While previous approaches mainly focus on Renyi entanglement quantities by using a finite number of replicas, our work directly addresses the von Neumann entropy, which is provably related to the decoding transition. As a by-product, this work also features an example where the von Neumann entropy has fundamentally different physics from other integer Renyi indices. 
In contrast to previous works on QECCs in gapless systems which mainly focus on code distances \cite{bentsen2023approximate,Kim_2017,Brandao_2019,yi2023complexity}, our results involve the decoding threshold for {\it extensive} noise. Furthermore, our results are applicable to generic CFTs as opposed to specific models. 

\textit{CFT code--} Let us start with properly defining the CFT code. Given a critical quantum spin chain with $n$ spins and Hamiltonian $H$, the low-energy physics is described by a CFT. Each low-energy eigenstate $|\phi_{\alpha}\rangle$ corresponds to a scaling operator $\phi_{\alpha}$ in the CFT due to the state-operator correspondence. The energy of the state is given by $E_{\alpha} = \frac{2\pi}{n} (\Delta_{\alpha}-c/12)$, where $\Delta_{\alpha}$ is the scaling dimension and $c$ is the central charge. As $n\rightarrow \infty$, there are infinite number of low-energy eigenstates whose energies are degenerate. The encoding circuit can be chosen as a MERA tensor network \cite{Vidal_2007,Pfeifer_2009,evenbly2013quantum,Ferris_2014,Kim_2017}, where the logical qubits are located at top layers. This connects CFT codes with tensor-network codes \cite{Farrelly_2021}, although here the tensors go beyond stabilizers.

In order to discuss the error-correcting properties, we specify the error model to be a finite-time evolution $e^{t\mathcal{L}}$ of a local Lindbladian $\mathcal{L}$, where $\mathcal{L}(\rho) = \sum_{i} \left[L_i \rho L^{\dagger}_i-\frac{1}{2}\{L^{\dagger}_i L_i,\rho\}\right]$ and $L_i$ are local jump operators. More concretely, we consider two types of noise, dephasing and flagged dephasing. For both error models, the noise channel $\mathcal{N} = \otimes_{j} \mathcal{N}^{[j]}$ is the product of local channels $\mathcal{N}^{[j]}$ acting on a single spin $j$.
A dephasing channel of strength $p$ is defined by
\begin{equation}
\label{eq:noise_unflagged}
    \mathcal{N}^{[j]}_{p,\alpha}(\rho) = \left(1-\frac{p}{2}\right) \rho + \frac{p}{2} \sigma^{[j]}_{\alpha} \rho \sigma^{[j]}_{\alpha},
\end{equation}
where $\sigma^{[j]}_{\alpha}$ is the Pauli operator acting on site $j$ and $\alpha=x,y,z$. In terms of Lindbladian, this corresponds to $n$ jump operators $L_i = \sigma^{[i]}_{\alpha}/\sqrt{2}$ and evolution time $t=-\log(1-p)$. A flagged dephasing channel (also known as heralded noise in quantum optics \cite{Brida_2011}) of strength $p$ is defined by
\begin{equation}
\label{eq:noise_flageed}
    \mathcal{N}^{[j]}_{p,\alpha;\mathrm{F}}(\rho) = \left(1-p\right) \rho \otimes |0_F\rangle \langle 0_F|+ p  \mathcal{N}^{[j]}_{1,\alpha}(\rho) \otimes |1_F\rangle \langle 1_F|,
\end{equation}
where we have introduced a ``flag" qubit $F$ at each site $j$. The flagged dephasing means that each site has a probability $p$ to undergo a complete dephasing, but we know which sites are subject to error by measuring the flag qubit. The flagged noise has $2n$ jump operators, $L_{2i-1} = I\otimes \sigma^{[i]}_{+}/\sqrt{2}, L_{2i} = \sigma^{[i]}_{\alpha} \otimes \sigma^{[i]}_{+}/\sqrt{2}$. 

We say that the noise is correctable if there exists a decoding channel $\mathcal{D}$ at each size $n$ such that $ \lim_{n\rightarrow \infty} F(\mathcal{D} \circ \mathcal{N} (\rho),\rho) = 1 $ for any $\rho$ in $\mathcal{H}_{\mathrm{code}}$, where $F$ is the Uhlmann fidelity. The uniform dephasing is harder to correct than the flagged dephasing, as one can discard the flag to obtain the uniformly dephased state, $\tr_{\mathrm{F}} \mathcal{N}_{p,\alpha;\mathrm{F}}(\rho) = \mathcal{N}_{p,\alpha}(\rho)$. 

\begin{figure}
    \centering
    \includegraphics[width = 0.35\linewidth]{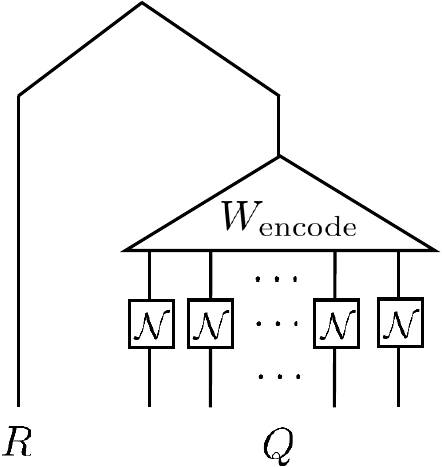}
    \includegraphics[width = 0.6\linewidth]{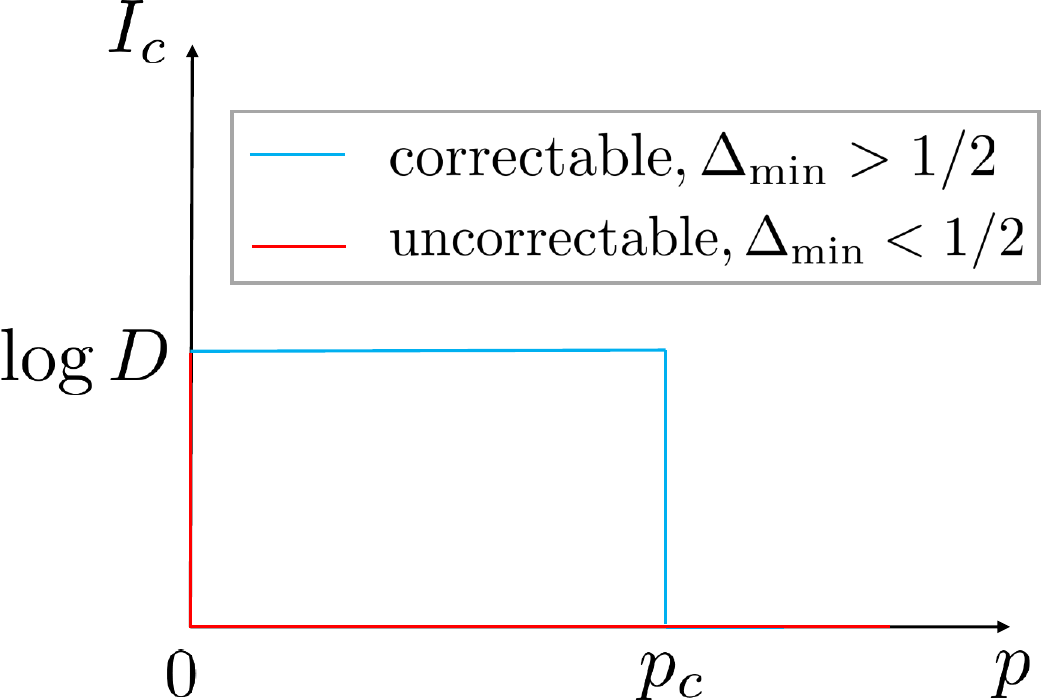}
    \caption{(Left) The coherent information of the CFT code is defined by $I_c = S_Q-S_{RQ}$ for the state $\rho_{RQ}$ in the figure, where $W_{\mathrm{encode}}$ is an encoding isometry which outputs states in the CFT low-energy subspace and $\mathcal{N}$ denote noise channels acting on all physical sites. (Right) Coherent information of correctable (blue) and uncorrectable (red) noise for the CFT code in the thermodynamic limit $n\rightarrow\infty$, where $D$ is the dimension of the logical subspace. The correctability condition is $\Delta_{\min}>1/2$, where $\Delta_{\min}$ is the smallest scaling dimension in the fusion algebra of the noise channel jump operators.} 
    \label{fig:Ic-TN}
\end{figure}

\textit{Error correcting condition.--} In order to characterize the recoverability of a QEC, Ref.~\cite{Schumacher_1996} introduces the entanglement fidelity for a decoding channel $\mathcal{D}$,
    $F_e = \langle \psi_{RQ}|\mathcal{D}\circ \mathcal{N}(|\psi_{RQ}\rangle\langle\psi_{RQ}|)|\psi_{RQ}\rangle$,
where 
\begin{equation}
    |\psi_{RQ}\rangle = \frac{1}{\sqrt{D}} \sum_{\alpha=1}^D  |\alpha\rangle_R|\phi_{\alpha}\rangle_Q
\end{equation}
is the maximal-entangled state between the reference $R$ and the code subspace of the physical qubits $Q$. 
The fidelity between any logical state and the recovered state is lower bound by $F_e$. Let $\rho_{RQ} = \mathcal{N}(|\psi_{RQ}\rangle\langle\psi_{RQ}|)$ (see Fig.~\ref{fig:Ic-TN}), one can further define the coherent information 
\begin{equation}
    I_c = S_Q-S_{RQ},
\end{equation}
where $S_{A} := -\tr(\rho_A\log \rho_A)$. It has been shown that if the coherent information is close to its maximal value, that is, $I_c = \log D - \epsilon$, then there exists a decoding channel $\mathcal{D}$ which achieves entanglement fidelity $F_e > 1-2\sqrt{\epsilon}$ \footnote{There is also an upper bound of $F_e$ by $1-F_e>O(\epsilon)$ due to monotonicity of coherent information and Fannes inequality.}. A QEC can perfectly correct the error if and only if $I_c = \log D$. If $\epsilon>0$, then the code is an approximate QEC.
A CFT code can approximately correct any noise that only acts on one lattice site. The reason is that the jump operator $L_i$ on site $i$ connects different codeword states with an amplitude \cite{CARDY1986186,Zou_2020conformal}
\begin{equation}
\label{eq:OPE}
    \langle \phi_{\beta}|L_i|\phi_{\alpha}\rangle = \left(\frac{2\pi}{n}\right)^{\Delta} C_{\alpha\beta L},
\end{equation}
where $\Delta$ is the scaling dimension of the operator $L_i$ and $C_{\alpha\beta L}$ is the operator product expansion (OPE) coefficient that is independent of $n$. This violation of the Knill-Laflamme condition implies that the best entanglement infidelity decreases to zero polynomially in $n$ \cite{yi2023complexity}.  

Below, we consider the CFT code under noise that uniformly acts on all sites, which is much harder to treat than the single-site noise. The coherent information $I_c(p,n)$ depends on the noise rate and the system size. The noise is correctable at error rate $p$ if and only if
\begin{equation}
\label{eq:decodability_Ic}
    \lim_{n\rightarrow \infty} I_c(p,n) = \log D ~(\mathrm{correctable~condition}).
\end{equation}
Given a noise model, such as flagged dephasing Eq.~\eqref{eq:noise_flageed} or unflagged dephasing Eq.~\eqref{eq:noise_unflagged}, the code is said to have a threshold $p_c$ if all noise channels with $p<p_c$ are correctable.

\textit{Perturbative expansion of coherent information--} In order to derive our main result, we make use of a perturbative expansion of coherent information together with the scaling hypothesis. The derivation is sketched below. 

The crucial assumption is the scaling hypothesis, that is, the coherent information has a scaling collapse of the form 
\begin{equation}
\label{eq:scaling}
    I_c(p,n) = f(pn^{\nu})
\end{equation}
as $n\rightarrow \infty$ and $p\rightarrow 0$, where $\nu$ is a constant that is analogous to the critical exponent. The same scaling form, in particular the same exponent $\nu$, is assumed to hold for any scaling of $p$ with respect to $n$ as long as $p\rightarrow 0$. The underlying physical intuition is to treat the noisy channel as a perturbation which flows towards or flows away from the no-noise fixed point under the renormalization group flow. Furthermore, the scaling form is also confirmed numerically, as we demonstrate later. The scaling function $f$ is monotonically decreasing and satisfies $f(0)=\log D$ and $f(\infty) < \log D$. For dephasing noise, one can easily show $f(\infty) = 0$. Therefore, the sign of $\nu$ determines the error correctability in the thermodynamic limit.
If $\nu>0$, then $I_c(p,n)\rightarrow 0$ in the thermodynamic limit for arbitrary $p>0$, so $p_c = 0$. If $\nu<0$, then the correctability condition in Eq.~\eqref{eq:decodability_Ic} is satisfied for small enough $p$, and thus the error has a finite threshold. 

Next, we perform a perturbation theory on coherent information assuming that $p=o(1/n)$. The quantum channel can be approximated by
\begin{equation}
    \mathcal{N}_p(\rho) = \rho + \frac{p}{2} \mathcal{L}(\rho) + O(p^2),
\end{equation}
where $\mathcal{L}$ is the Lindbladian.
The state $\rho_{RQ}(p) = \mathcal{N}(|\psi_{RQ}\rangle\langle\psi_{RQ}|)$ can be expanded to first order in $p$. Expanding $S_Q$ and $S_{QR}$ to the first order (see details in appendix), we obtain
\begin{equation}
\label{eq:expansion}
    I_c(p,n) = \log D + b(n) p\log p + O(p),
\end{equation}
where
\begin{equation}
\label{eq:b}
    b(n) = \frac{1}{D^2}\sum_{i=1}^n[D\tr(L_i P L^{\dagger}_i P)-\tr(L_i P)\tr(L^{\dagger}_i P)]
\end{equation}
and $P = \sum_{\alpha} |\phi_{\alpha}\rangle \langle \phi_{\alpha}|$ is the projector onto the code subspace. The noise is correctable only if $\lim_{n\rightarrow \infty} b(n) = 0$. It is worth noting that $b(n)\geq 0$ and the equality holds if and only if $PL_i P \propto P$, which is part of the Knill-Laflamme conditions for a single-qubit error. Higher-order perturbations can reproduce Knill-Laflamme conditions with multiqubit errors.

Finally, using Eq.~\eqref{eq:OPE} we obtain $b(n)\propto n^{1-2\Delta}$, where $\Delta$ is the scaling dimension of the jump operator. The only way that $b(n)\propto n^{1-2\Delta}$ can be compatible with the scaling form $I_c(p,n) = f(pn^{\nu})$ is that
\begin{equation*}
    \nu = 1-2\Delta.
\end{equation*}
Crucially, the scaling hypothesis allows us to infer the exponent $\nu$ for $p=O(1)$ from the finite-order perturbation theory performed at $p=o(1/n)$. Thus, the error correctability condition $\nu>0$ holds at finite $p$ if and only if $\Delta>1/2$. 

Expanding to higher orders of perturbation theory, we find that correctability requires that all OPE between $L_i$ and itself contain operators with scaling dimension larger than $1/2$, excluding the identity operator. If the lowest operator with scaling dimension $\Delta_{\min}$ appears at fusion order $r$, then $\nu = (1-2\Delta_{\min})/r$. We provide a calculation to the second order and an argument for higher orders in the appendix.



In order to verify our result, we numerically compute coherent information for both unflagged (Eq.~\eqref{eq:noise_unflagged}) and flagged (Eq.~\eqref{eq:noise_flageed}) dephasing noise. We remark that two noise models should have the same $\nu$ according to our argument, because they contain the same jump operators that act on the physical system. For the unflagged case, we are restricted to small system sizes due to exponentially growing simulation complexity. For the flagged case, we can access much larger system sizes by sampling over measurement trajectories, which is a trick first explored in Ref.~\cite{Bao_2020} and detailed in the appendix.


\begin{figure*}[t]
    \centering
    \includegraphics[width=0.33\linewidth]{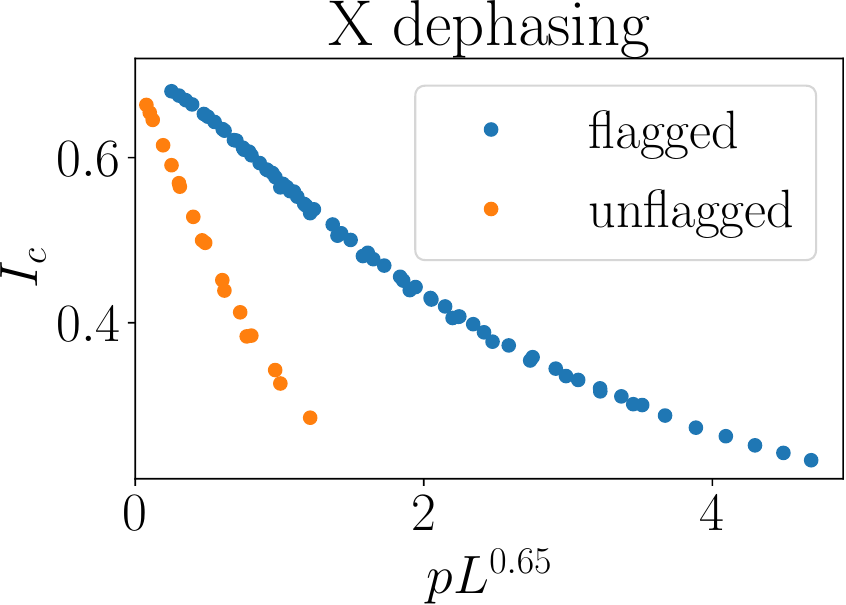}
    \includegraphics[width=0.33\linewidth]{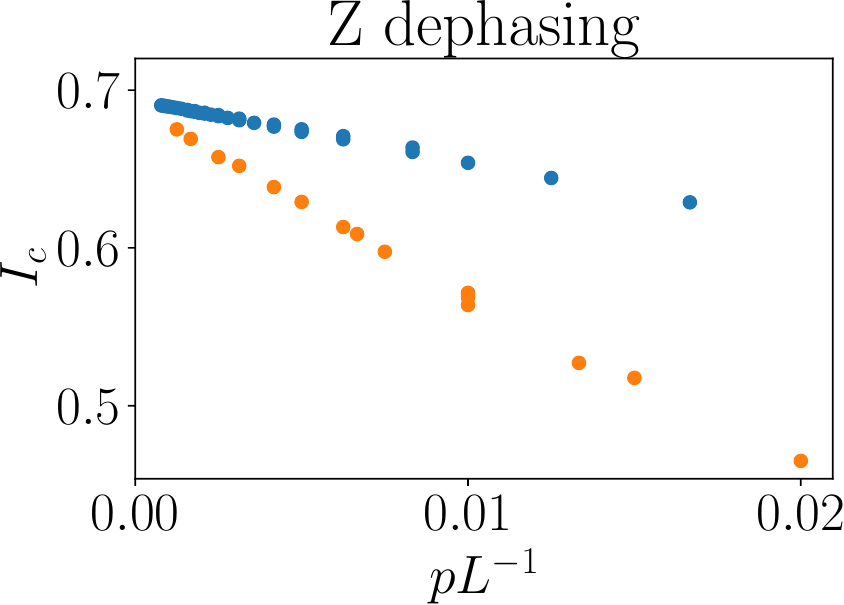}
    \includegraphics[width=0.31\linewidth]{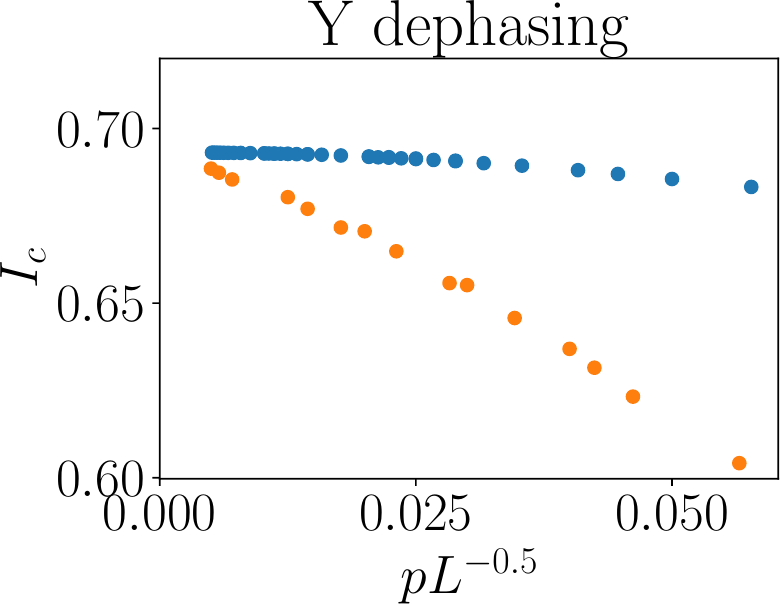}
    \caption{Coherent information of the Ising CFT code under flagged and unflagged dephasing with $p\leq 0.2$. The data collapses show that the exponents $\nu$ are approximately $0.65,-1.0$ and $-0.5$ for $X,Z$ and $Y$ dephasings, irrespective of whether the dephasing is flagged or not.}
    \label{fig:CI_smallp}
\end{figure*}

\textit{Example: Ising CFT code.--} As an example, let us consider the Ising CFT realized by the transverse field Ising model $H = -\sum_{i} (\sigma^{[i]}_x \sigma^{[i]}_x + g \sigma^{[i]}_z)$ at $g=g_c=1$. It has three primary operators labelled by $I,\sigma,\varepsilon$ with scaling dimensions $\Delta_I = 0, \Delta_{\sigma}=1/8, \Delta_{\varepsilon} = 1$ \cite{BELAVIN1984333}. The code subspace can be minimally chosen to be spanned by the ground state $|I\rangle$ and the second excited state $|\varepsilon\rangle$. The three single-site Pauli operators $\sigma_{x},\sigma_{y},\sigma_{z}$ corresponds to $\sigma, \partial\sigma,\varepsilon$ in the CFT \cite{Zou_2020conformal}. We can read off the exponent $\nu$ for the $X$, and $Z$ dephasing $\nu_x = 1-2\Delta_{\sigma} = 0.75$ and $\nu_z = 1-2\Delta_{\varepsilon} = -1$. For $Y$ dephasing, we can use second order perturbation theory to show that $\nu_y = (1-2\Delta_{\varepsilon})/2 = -0.5$. The underlying reason is that the OPE $\partial\sigma\times \partial\sigma $ contains an operator $\varepsilon$ with $\Delta_{\varepsilon}<\Delta_{\partial\sigma}$, and further fusion does not produce a lower one than $\varepsilon$. Thus our theory predicts that the $Y$ and $Z$ dephasing have a finite threshold and $X$ dephasing has zero threshold.

We compute the coherent information for the unflagged dephasing up to $n=16$ \footnote{We use MPO to store the purified state $|\psi\rangle_{RQE}$ and coarse-grain $Q$ and $E$ separately. The effective physical dimension must increase exponentially with subsystem sizes, so we have limited system size $n\leq 16$. } and flagged dephasing up to $n=128$ \footnote{For the sampling algorithm we use 56000 samples.} with $p\leq 0.2$, see Fig.~\ref{fig:CI_smallp}. The scaling collapses fit best with $\nu_x \approx 0.65,\nu_y \approx -0.5, \nu_z\approx -1.0$, which are close to the theoretical values.

\begin{figure}[htbp]
    \centering
    \includegraphics[width=0.99\linewidth]{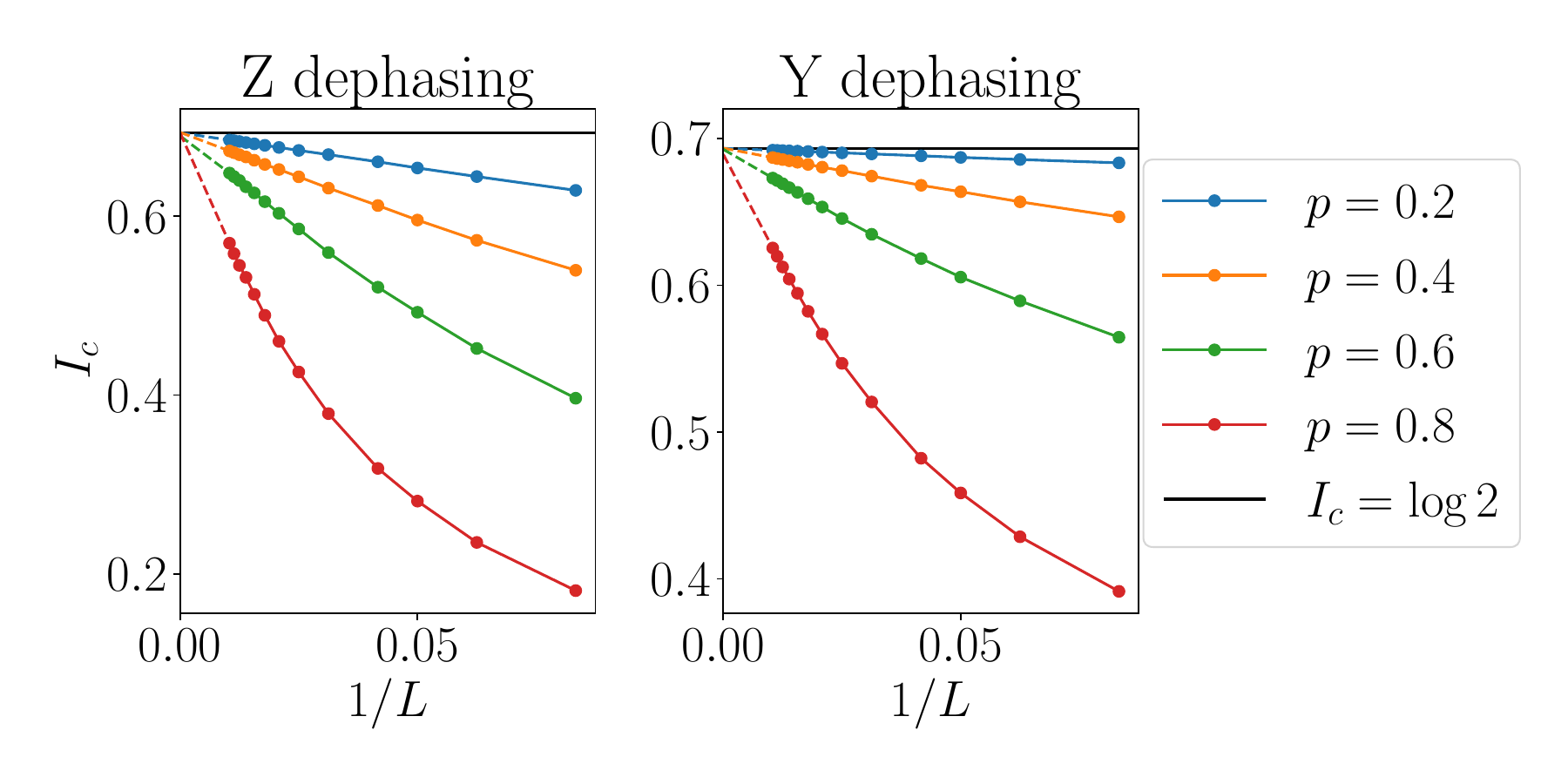}
    \caption{Coherent information of the Ising CFT code under $Z$ (left) and $Y$ (right) flagged dephasing with $0.2\leq p\leq 0.8$. As $L\rightarrow\infty$, the coherent information extrapolates to $\log 2$, indicating the correctability of the error.}
    \label{fig:CI_flagged}
\end{figure}

In order to obtain the threshold, we compute $I_c$ for flagged noise at large $p$ and find that the threshold $p=1$ for both $Y$ and $Z$ dephasing. In Fig.~\ref{fig:CI_flagged} we show the extrapolation of $I_c$ to the thermodynamic limit and show that $I_c = \log 2$ up to $p\leq 0.8$. We also perform a scaling collapse near $p=1$ to confirm that it is indeed an unstable RG fixed point, which indicates that $p=1$ is the threshold. For unflagged dephasing, the threshold can be estimated by the crossing point of the curves $I_c(p)$ of different $n$. Our small-size simulation suggests that the threshold is also $p_y=p_z = 1$, see appendix for details.

We see that the correctable and uncorrectable dephasing of the Ising CFT code is identical to the repetition code. It is not unexpected since the repetition code can be realized by the spontaneous symmetry broken phase $0<g<1$ of TFIM. 
However, as opposed to the repetition code where the number of protected logical qubits is $k=O(1)$, the CFT code can protect more logical information as we increase $n$, as we demonstrate below.

\textit{Dimension of logical subspace.--} The low-energy subspace is infinite-dimensional as $n\rightarrow\infty$, as each primary operator has infinitely many descendants. It is then natural to ask how large the code subspace can be chosen to maintain a finite threshold for correctable dephasing noise with $\Delta>1/2$. Firstly, as $ b(n)\propto n^{1-2\Delta}$ in Eq.~\eqref{eq:b} for any finite-dimensional code subspace, we can store at least constant number of qubits.  Below, we further argue that the dimension of code subspace can scale at least as $\mathrm{polylog}(n)$, thus $k \geq  \Omega(\log \log n)$. Let us choose the code subspace to be spanned by a primary state $|\phi\rangle$ and its global descendants $|\partial^m \bar{\partial}^{\bar{m}} \phi\rangle$, where $m$ and $\bar{m}$ are natural numbers smaller than a cutoff $M$. Then the projector to the code subspace is
\begin{equation}
    P = \sum_{m,\bar{m}=0}^{M-1} |\partial^m \bar{\partial}^{\bar{m}} \phi\rangle\langle\partial^m \bar{\partial}^{\bar{m}} \phi|
\end{equation}
As we show in the appendix, if $M=\mathrm{polylog}(n)$, then $\lim_{n\rightarrow \infty} b(n)\rightarrow 0$ given any $\Delta > 1/2$. The total dimension of the code subspace $2^k = M^2$, which means $k \propto \log(M) = O(\log\log n)$. 
Thus, we establish that the number of logical qubits increases with the number of physical qubits. In principle, one can encode more logical qubits by including Virasoro descendants in the code subspace. By including all descendants below scaling dimension $M = \mathrm{polylog}(n)$, one may reach $k=O(\mathrm{polylog}(n))$ by the Cardy formula \cite{CARDY1986186}. We leave it to future work to investigate its error protection properties.



\textit{Covariant CFT code.--} Let us consider a particular case of the CFT code where the CFT has a global $U(1)$ symmetry. Then there exists a conserved charge density which has scaling dimension $\Delta = 1>1/2$. Thus the dephasing noise on the charge density is a correctable error. Furthermore, $\nu = 1-2\Delta=-1$ in Eq.~\eqref{eq:scaling}, which implies that the drop of coherent information $\epsilon := \log D - I_c \propto n^{-1}$. Together with the result of Ref.~\cite{schumacher2001approximate} that the entanglement fidelity $F_e> 1-2\sqrt{\epsilon}$, we have $1-F_e \leq O(n^{-1/2})$.  The same conclusion holds for arbitrary dimensions. On the other hand, a covariant code has fundamental limitations on quantum coherence~\cite{Faist2020continuous,Liu_2023approximate, tajima2022universallimitationquantuminformation,tajima2022universaltradeoffstructuresymmetry}. Specifically, for any covariant code with $U(1)$ symmetry, Refs.~\cite{Liu_2023approximate, tajima2022universallimitationquantuminformation} have shown that $ 1-F_e \geq O(n^{-1/2})$. Thus, the only possibility is that the $U(1)$-symmetric CFT code satisfies $1-F_e = O(n^{-1/2})$, which saturates the covariant code bound above.


\textit{Discussion.--} Our result can be readily applied to depolarization noise, where all Pauli errors can happen. The CFT code can correct depolarization noise at a finite threshold if and only if all scaling dimensions $\Delta>1/2$. Such a CFT does not exist for Virasoro minimal models and all $c=1$ free boson models. CFTs that satisfy this condition have $c>1$ and may only be realized beyond spin-$1/2$ chains and $2-$local interactions \cite{latorre2024cd}. One such example is the monster CFT with the lowest primary $\Delta=4$, which has been conjectured to be dual to a bulk quantum gravity theory \cite{witten2007threedimensional}. 
Another example that has $\Delta_{\min}>1/2$ is the $SU(N)_1$ WZW model for $N\geq 3$, where the simplest case $N=3$ can be realized by a spin-1 chain \cite{Mashiko_2021}. 
One interesting feature of these CFT codes is that they may have a finite threshold in one spatial dimension, which cannot be achieved by the ground state subspace of any gapped local Hamiltonian.


In terms of characterization of mixed-state phases, one of our technical contributions is the perturbative calculation of coherent information without applying replica trick, while prior works mostly focus on finite replica calculations. In fact, the Renyi coherent information may give incorrect predictions of the code properties. For example, it gives a wrong decoding threshold for the toric code \cite{bao2023mixed}. This work has given another example which shows that working with integer Renyi index predicts wrong error-correctability of infinitesimally small noise strength. For the Ising CFT code, uniform $Z$ dephasing can be corrected at a finite threshold but the Renyi coherent information drops as long as $p\neq 0$, see appendix for details. 

Our work opens up several future directions. Firstly, even though we have shown that some errors can be corrected for CFT code, it is still nontrivial to construct an explicit decoder, an important ingredient of QECC. We expect that good decoders can be obtained through real-space renormalization group \cite{sang2023mixedstate, Furuya_2022, Furuya_2022_2, Kuwahara_2024} or classical optimization techniques \cite{Fletcher_2007}. Secondly, we can consider logical operations of CFT codes and see if they can be performed fault-tolerantly. Thirdly, it is interesting to consider the bulk dual in AdS/CFT with boundary dephasing \cite{Mintun_2015,Milekhin_2021}, which may inspire the solution of the two questions above.

\textit{Acknowledgements.--} We thank Charles Cao, Yu-Hsueh Chen, Tarun Grover, Yangrui Hu, Isaac Kim, Yaodong Li, David Lin, Peter Lu, Ruochen Ma, Alexey Milekhin, Daniel Parker, Bowen Shi,  Guifre Vidal, Sisi Zhou, Tianci Zhou and Zheng Zhou for helpful discussions. Y.Z. in particular thanks Patrick Hayden for his insight that leads this work. This work was supported by the Perimeter Institute for Theoretical Physics (PI) and the Natural Sciences and Engineering Research Council of Canada (NSERC).  Research at PI is supported in part by the Government of Canada through the Department of Innovation, Science and Economic Development Canada and by the Province of Ontario through the Ministry of Colleges and Universities. This research was supported in part by grant NSF PHY-2309135 to the Kavli Institute for Theoretical Physics (KITP).

\bibliography{ref}

\onecolumngrid

\section*{Flagged dephasing and unflagged dephasing}
\subsection{Coherent information}
In order to compute the coherent information, let us start with an unravelling of the flagged dephasing. At each site there is a probability $1-p$ that we don't measure the state, and probability $p$ that we measure the state in the $\sigma_{\alpha}$ basis, with a Born probability $q_{\pm} = \mathrm{tr}(\rho_{RQ} \frac{I\pm \sigma_{\alpha}}{2})$ to project onto the $|\pm \alpha \rangle$ state. A crucial observation is the output states of the above three cases are mutually orthogonal. This is because, firstly, if the measured sites are different, then the flags are orthogonal, and secondly, if the measurement outcomes are different, then the measured qubits are orthogonal. Thus, $\rho_{RQF} = \mathcal{N}_{p,\alpha;F}(|\psi_{RQ}\rangle\langle\psi_{RQ}|)$ can be unravelled as
\begin{equation}
    \rho_{RQF} = \sum_{\mathbf{m}} p_{\mathbf{m}} |\psi^{\mathbf{m}}_{RQ}\rangle \langle \psi^{\mathbf{m}}_{RQ}| \otimes |\mathbf{m}\rangle \langle \mathbf{m}|.
\end{equation}
Note that here $|\mathbf{m}\rangle$ records not only which sites are measured but also the outcome state in the measured sites. More precisely, denote $m=\pm$ as measurement outcome and $m=0$ as no measurement, and $P^{[j]}_{\pm} = \frac{I\pm \sigma^{[j]}_{\alpha}}{2}$, $P^{[j]}_0 = I$ then
\begin{eqnarray}
    |\psi^{\mathbf{m}}_{RQ}\rangle = \sqrt{p_{\mathbf{m}}} \left(\prod_{j} P^{[j]}_{\mathbf{m}_j} \right)|\psi_{RQ}\rangle. 
\end{eqnarray}
It follows that
\begin{eqnarray}
    S(\rho_{RQF}) &=& -\sum_{\mathbf{m}} p_{\mathbf{m}} \log p_{\mathbf{m}} \\
    S(\rho_{QF}) &=& -\sum_{\mathbf{m}} p_{\mathbf{m}} \log p_{\mathbf{m}} + \sum_{\mathbf{m}} p_{\mathbf{m}} S_{Q}(|\psi^{\mathbf{m}}_{RQ}\rangle).
\end{eqnarray}
Note that $S_Q = S_R$ for a pure state $|\psi^{\mathbf{m}}_{RQ}\rangle$. The coherent information $I_c = S(\rho_{QF}) - S(\rho_{QRF})$ is then
\begin{equation}
    I_c = \mathbb{E}_{\mathbf{m}} S_R(|\psi^{\mathbf{m}}_{RQ}\rangle),
\end{equation}
where $\mathbb{E}_{\mathbf{m}}$ is the average with respect to the joint probability distribution $p_{\mathbf{m}}$ of the flags and measurement outcomes.
\subsection{Threshold of flagged dephasing}
We can calculate the coherent information of the Ising CFT code under flagged dephasing using MPS techniques. We first use DMRG to diagonalize the low-energy eigenstates in the code subspace. We then construct the maximal entangled state $\psi_{RQ}$ as an MPS by concatenating the MPS of the eigenstates. We compress the MPS to lower bond dimension by allowing a small amount $10^{-8}$ of truncation error. We then perform measurement on $Q$ and obtain a sample $|\psi^{\mathbf{m}}_{RQ}\rangle$ of post-measurement states as an MPS, from which we can read off the entanglement entropy $S_R$. Finally we average over $S_R$ to obtain the coherent information. The results for the flagged dephasing up to $n=128$ and $p\leq 0.8$ are shown in the main text. Here we provide additional data for the flagged $Y$ dephasing and the data near $p\approx 1$.

\begin{figure}[htbp]
    \centering
    \includegraphics[width = 0.49\linewidth]{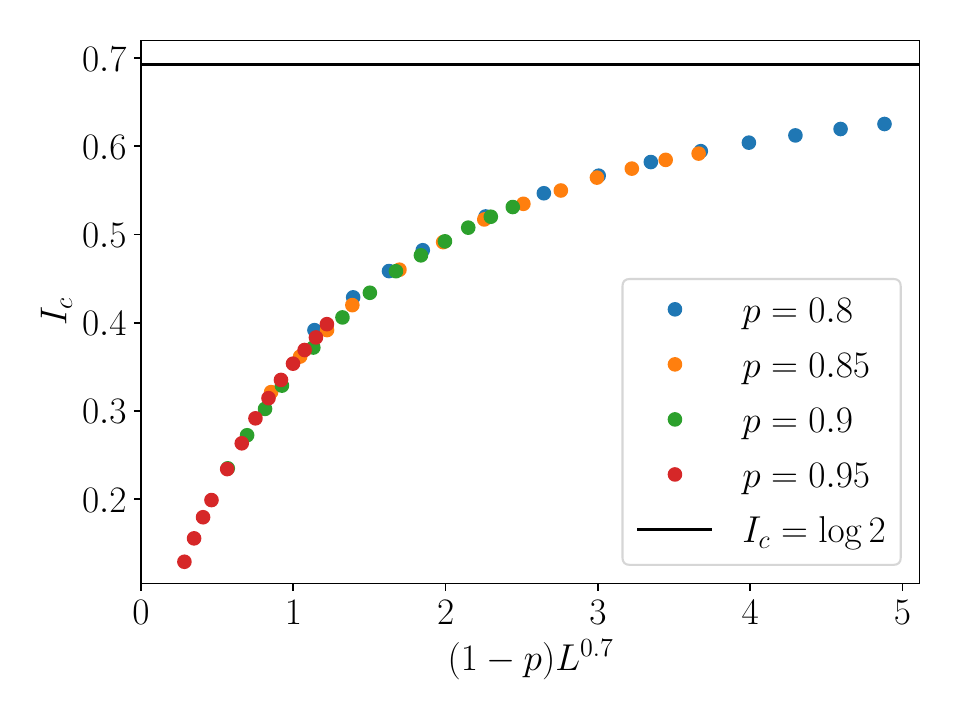}
    \includegraphics[width = 0.49\linewidth]{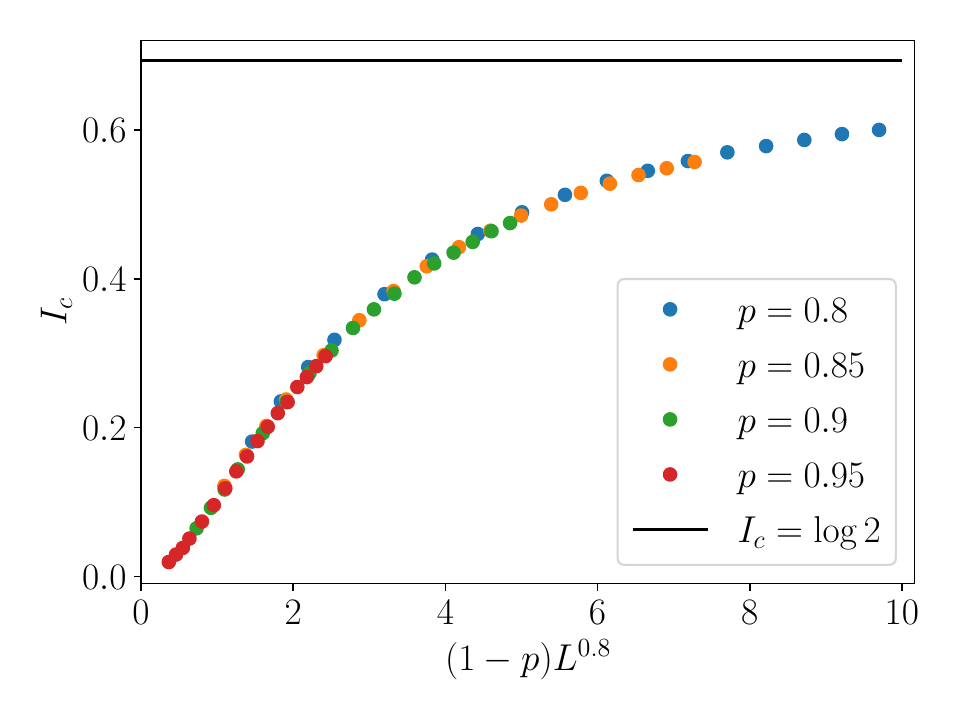}
    \caption{Flagged $Y$ (left) and $Z$ (right) dephasing coherent information of the Ising CFT code near $p=1$.}
    \label{fig:FCI_Y}
\end{figure}

For both $Y$ and $Z$ flagged dephasing, we can perform a scaling collapse near $p=1$ to obtain a scaling form $I_c((1-p)L^{\nu'})$ with $\nu'>0$. This indicates that $p=1$ is an unstable RG fixed point which flows to $p=0$. Thus, for both $Y$ and $Z$ flagged dephasing, the decoding threshold is $p_c = 1$.

\subsection{Threshold of unflagged dephasing}
We provide additional evidence tha that the threshold for unflagged $Y$ and $Z$ dephasing for the Ising CFT code is also $p_c=1$. We plot $I_c(p)$ with system sizes $7\leq  n \leq 11$ and find that the curves only intersect at $p=0$ and $p=1$. This suggests that the threshold is $p_c=1$ but the evidence is not conclusive due to very small system sizes.

\begin{figure}[htbp]
    \centering
    \includegraphics[width = 0.49\linewidth]{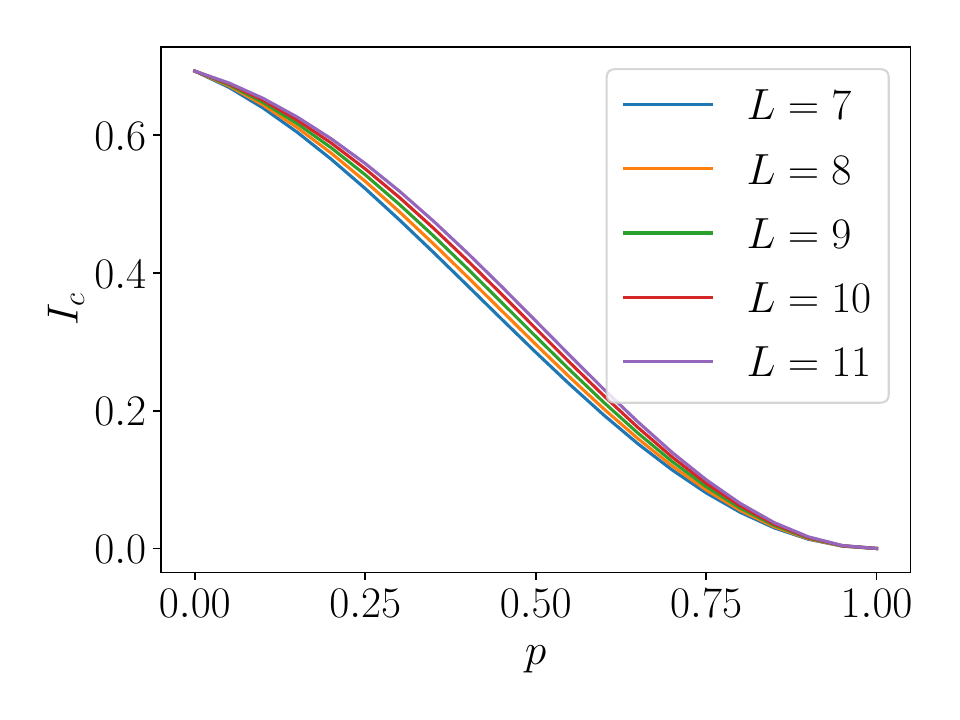}
    \includegraphics[width = 0.49\linewidth]{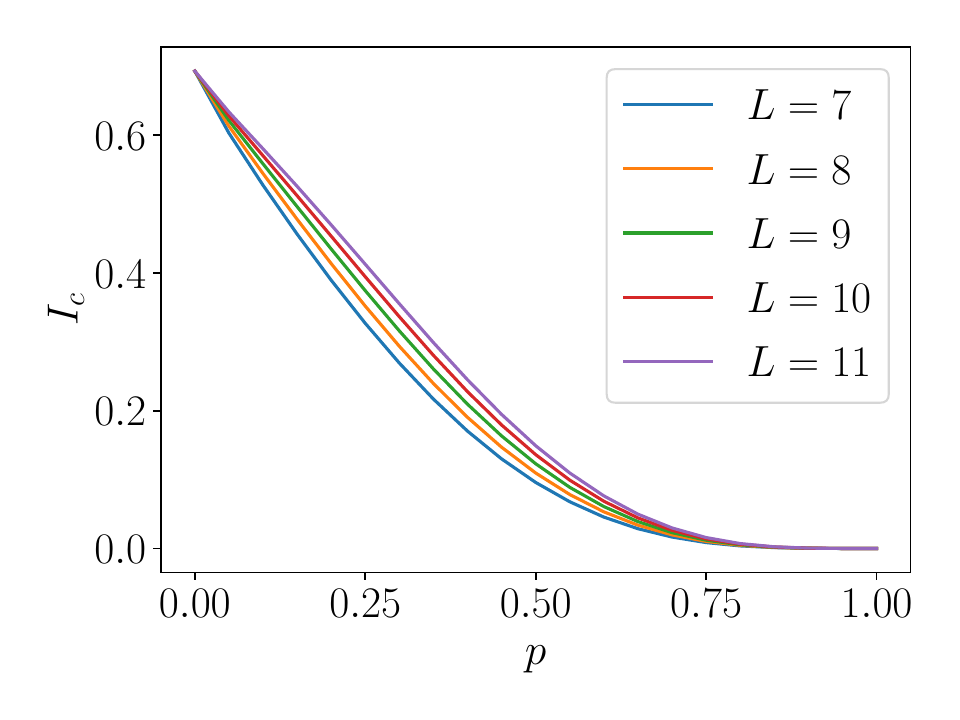}
    \caption{$Y$ (left) and $Z$ (right) dephasing coherent information of the Ising CFT code. The only intersections of the curves are at $p=0$ and $p=1$, indicating a threshold $p_c=1$ in the thermodynamic limit.}
    \label{fig:FCI_Y}
\end{figure}

\section*{Perturbation of coherent information}
In this section we consider the change of coherent information under an infinitesimal evolution with the Lindbladian.
\subsection{First order perturbation}
Under the first-order perturbation, we can approximate
\begin{equation}
    \rho_{RQ}(p) = |\psi\rangle\langle \psi| + p \sum_{i=1}^n \left(L_i|\psi\rangle\langle \psi| L^{\dagger}_i - \frac{1}{2} L_i^{\dagger} L_i |\psi\rangle\langle \psi| - \frac{1}{2}  |\psi\rangle\langle \psi| L_i^{\dagger} L_i\right) + O(p^2),
\end{equation}
where
\begin{equation}
\label{eq:SQR1}
    |\psi\rangle = \frac{1}{\sqrt{D}}\sum_{\alpha=1}^D |\alpha\rangle_R |\phi_{\alpha}\rangle_Q. 
\end{equation}
We will look at the eigenvalues of the density matrix $\rho_{RQ}(p)$. It is clear that there are at most $n+1$ nonzero eigenvalues, which are  $\lambda_0 = 1-p\nu_0$ and $\lambda_i = p\nu_i$ to the first order in $p$. Since the first eigenvalue is non-degenerate, we can use first-order perturbation theory on the eigenvalue to obtain
\begin{equation}
\label{eq:SQR2}
    -\nu_0 = \sum_{i}\left(|\langle \psi |L_i|\psi\rangle|^2 - \langle \psi|L^{\dagger}_i L_i| \psi \rangle\right)
\end{equation}
Now we can compute the entropy $S_{RQ}$ to be
\begin{eqnarray}
    S_{RQ} &=& -\sum_{i}\lambda_i \log \lambda_i \\
    &=& -(1-p\nu_0) \log (1-p\nu_0) - \sum_{i} p\nu_i \log (p\nu_i) \\
    &=& -\left(\sum_{i} \nu_i\right) p\log p + \left(\nu_0 - \sum_i \nu_i \log \nu_i\right)p + O(p^2).\\
    & = & \nu_0 p\log p + \left(\nu_0 - \sum_i \nu_i \log \nu_i\right)p + O(p^2) \label{eq:SQR3}
\end{eqnarray}
where we have used the fact that $\tr\rho_{RQ}(p) = 1$ in the last step.
Now we turn to the state $\rho_Q(p)=\tr_R \rho_{RQ}(p)$ and consider its eigenvalues. To the first order in $p$,
\begin{equation}
    \rho_{Q}(p) = \sum_{\alpha=1}^D \frac{1}{D}|\phi_{\alpha}\rangle\langle \phi_{\alpha}| + \frac{p}{D} \sum_{\alpha=1}^D \sum_{i=1}^n \left(L_i|\phi_{\alpha}\rangle\langle \phi_{\alpha}| L^{\dagger}_i - \frac{1}{2} L_i^{\dagger} L_i |\phi_{\alpha}\rangle\langle \phi_{\alpha}| - \frac{1}{2}  |\phi_{\alpha}\rangle\langle \phi_{\alpha}| L_i^{\dagger} L_i\right) + O(p^2).
\end{equation}
The first $D$ eigenvalues are then $(1/D-p\alpha_1,1/D-p\alpha_2,\cdots 1/D-p\alpha_D)$ and the rest are proportional to $p$ with at most $Dn$ of them, $(p\beta_1,\cdots p\beta_{Dn})$. Using the degenerate perturbation theory on the subspace spanned by the first $D$ eigenvalues, we can show that $-\alpha_k$'s are the eigenvalues of a $D\times D$ matrix $V$ whose components are
\begin{eqnarray}
\label{eq:SQ1}
    V_{kk} &=& \frac{1}{D}\left(\sum_{i,\alpha} \langle \phi_k|L_i |\phi_{\alpha}\rangle \langle \phi_{\alpha}|L^{\dagger}_i|\phi_k\rangle - \sum_{i} \langle \phi_k|L^{\dagger}_i L_i|\phi_k \rangle\right)\\
    V_{kl} &=& \frac{1}{D}\left(\sum_{i,\alpha} \langle \phi_k|L_i |\phi_{\alpha}\rangle \langle \phi_{\alpha}|L^{\dagger}_i|\phi_l\rangle - \sum_{i} \mathrm{Re}\langle \phi_k|L^{\dagger}_i L_i|\phi_l\rangle\right) ~~~~~(k\neq l)
\end{eqnarray}
In terms of the eigenvalues of $\rho_Q$, we can express
\begin{equation}
\label{eq:SQ2}
    S_{Q} = \log D  + \left(\sum_{k=1}^D \alpha_k\right) p\log p + \left( (1-\log D)\left(\sum_{k=1}^D \alpha_k \right) - \sum_{i}\beta_i \log \beta_i \right) p +O(p^2)
\end{equation}
Since the sum of eigenvalues equals the trace,
\begin{equation}
\label{eq:SQ3}
    \sum_{k=1}^D \alpha_k = -\sum_{k} V_{kk}.
\end{equation}
Combining Eqs.~\eqref{eq:SQR1}, \eqref{eq:SQR2}, \eqref{eq:SQR3}, \eqref{eq:SQ1}, \eqref{eq:SQ2}, \eqref{eq:SQ3} we obtain
\begin{equation}
\label{eq:appIc}
    I_c = S_Q-S_{QR} = \log D + b p\log p + O(p)
\end{equation}
where
\begin{equation}
    b = \frac{1}{D^2} \sum_{i,\alpha,\beta} (D|\langle \phi_{\beta}|L_i|\phi_{\alpha}\rangle|^2-\langle \phi_{\alpha}|L_i|\phi_{\alpha}\rangle \langle \phi_{\beta}|L^{\dagger}_i |\phi_{\beta}\rangle).
\end{equation}
More compactly we can write
\begin{equation}
\label{eq:b_compact}
    b = \frac{1}{D^2}\sum_{i=1}^n[D\tr(L_i P L^{\dagger}_i P)-\tr(L_i P)\tr(L^{\dagger}_i P)]
\end{equation}
where
\begin{equation}
    P = \sum_{\alpha=1}^D |\phi_{\alpha}\rangle\langle \phi_{\alpha}|
\end{equation}
is the projector onto code subspace.

We have shown in the main text that $b \propto n^{1-2\Delta}$, where $\Delta$ is the scaling dimension of the jump operator. If $\Delta<1/2$. then the leading correction term $bp\log p$ dominates Eq.~\eqref{eq:appIc} at small $p$ and gives $I_c<\log D$, thus the noise is uncorrectable. If $\Delta>1/2$, then the leading correction term $bp\log p \rightarrow 0$ as $n\rightarrow \infty$. We further argue below that the subleading correction $O(p)$ term in Eq.~\eqref{eq:appIc} gives the same exponent $1-2\Delta$ as $n\rightarrow \infty$. Using the scaling hypothesis $I_c = I_c(pn^{\nu})$ thus fixes $\nu=1-2\Delta$ and gives a finite threshold for the noise. 


\subsection{Second-order perturbation}
Before we do an explicit calculation, We will first intuitively argue how higher-order terms could potentially dominate the change of $I_c$ and how it affects the error correctability condition. Let us consider a block spin transformation of two spins $i$ and $i+1$ which are blocked into a four-dimensional local Hilbert space. Then for each site there is $O(p)$ probability where one error $L_i$ or $L_{i+1}$ happens and there is $O(p^2)$ probability where two errors $L_i L_{i+1}$ happen. Now $L_i L_{i+1}$ is a local operator and has its own scaling dimension $\Delta^{(2)}$. For a lattice realization of CFT, if $L_i$ corresponds to a scaling operator $O_{\Delta}$ with scaling dimension $\Delta$, then $L_i L_{i+1}$ corresponds to the lowest operator (except identity) in the operator product expansion $O_{\Delta} \times O_{\Delta} \rightarrow I + :O^2: + \cdots$. The normal-ordered operator $:O^2:$ has its own scaling dimension $\Delta^{(2)}$. If $\Delta^{(2)}<\Delta$, then the change of coherent information is dominated by this error for large system sizes, $\delta I_c \propto p^2 n^{1-2\Delta^{(2)}}\log p$. This then indicates that $\nu = (1-2\Delta^{(2)})/2$. More generally, within the self-OPE of the jump operator, if the lowest operator with scaling dimension $\Delta^{(r)}$ appears at the $r$-th order of fusion, then $\nu = (1-2\Delta^{(r)})/r$. In order to require the error to be correctable, we need all $\Delta^{(r)}>1/2$. This could happen if all of the CFT primary operators have scaling dimensions larger than $1/2$, or there is a symmetry of the jump operator $L$ that forbids it from fusing into primary operators with scaling dimension smaller than $1/2$. 

For an explicit example, consider the $Z$ operator the TFIM model which corresponds to the $I+\varepsilon$ operator in the CFT. Since we have the fusion rule $I\times \varepsilon = \varepsilon$ and $\varepsilon\times \varepsilon = I$, we will never encounter the $\sigma$ operator in the OPE, thus the lowest operator $\Delta^{(r)}= \Delta_{\varepsilon} = 1$ occurs at $r=1$. Thus $\nu = 1-2\Delta_{\varepsilon} = -1$. A more sophisticated example happens for the $Y$ dephasing, which corresponds to $Y = \partial_{t} \sigma$ in the Ising CFT. Two $Y$'s could fuse $\partial_{t} \sigma\times \partial_{t} \sigma = I + \varepsilon$ and three $Y$'s can only fuse into itself, $\partial_{t} \sigma\times \partial_{t} \sigma \times \partial_{t} \sigma = \partial_{t} \sigma$. Note that fusion into $\sigma$ is not possible due to time reversal symmetry. Thus, the lowest operator in the fusion appears at $r=2$, which is again the $\varepsilon$ operator. As a result $\nu = (1-2\Delta_{\varepsilon})/2 = -0.5$.

Now we perform a second order perturbation theory on the coherent information $I_c = S_Q - S_{QR}$ and see how the intuition above works. To start with, we can expand
\begin{equation}
\label{eq:Ic_expansion_2nd_order}
    I_c = \log D + bp \log p + b^{(2)}p^2 \log p + \mathrm{regular}
\end{equation}
and we will compute $b^{(2)}$ as a function of jump operators and system size. Let us consider the perturbed density matrix
\begin{equation}
\label{eq:expansion_2nd_order}
    \rho_{QR}(p) = \rho_{QR}(0) + p \mathcal{L}(\rho_{QR}(0)) + \frac{p^2}{2} \mathcal{L}(\mathcal{L}(\rho_{QR}(0))) + O(p^3)
\end{equation}
where $\rho_{QR}(0) = |\psi\rangle\langle \psi|$. The eigenvalues of $\rho_{QR}(p)$ can be parameterized as $1-p\nu_0 + p^2 \nu^{(2)}_0$ and $p\nu_i + p^2 \nu^{(2)}_i$. To this order, we can expand
\begin{equation}
    S_{QR} = \nu_0 p \log p - \nu^{(2)}_0 p^2\log p + \mathrm{regular.}
\end{equation}
where $\mathrm{regular}$ means terms analytic in $p$. We will see that  $\nu^{(2)}_0 \propto n^{1-2\Delta^{(2)}}$ using perturbation theory. There are two contributions, one coming from the second-order perturbation with the $p \mathcal{L}(\rho_{QR}(0))$ term and one coming from the first-order perturbation with the $\frac{p^2}{2} \mathcal{L}(\mathcal{L}(\rho_{QR}(0)))$ term. Note that $\rho_{QR}(0)$ has one eigenvalue $1$ and the rest are $0$. Denote for shorthand $\rho := \rho_{QR}(0), \delta \rho := \mathcal{L}(\rho_{QR}(0))$ and $\delta^2\rho := \mathcal{L}(\mathcal{L}(\rho_{QR}(0)))$ we have
\begin{equation}
    \nu^{(2)}_0 = \tr(\rho \delta\rho (1-\rho) \delta\rho) + \frac{1}{2}\tr(\rho \delta^2 \rho).
\end{equation}
For dephasing noise, we have $L^{\dagger}_i L_i = I$. This simplifies the expression but does not affect the conclusion. For this case we have
\begin{equation}
    \delta \rho = \sum_{i} \frac{L_i \rho L^{\dagger}_i-\rho}{2}, ~~ \delta^2 \rho = \sum_{i,j} \frac{L_i L_j \rho L^{\dagger}_j L^{\dagger}_i -\rho}{4}
\end{equation}
Now we can substitute the above equation to $\nu^{(2)}_0$ to obtain
\begin{equation}
    \nu^{(2)}_0 = \frac{1}{8}\sum_{i,j} \left( 2\langle L_i \rangle \langle L^{\dagger}_i L_j \rangle \langle L^{\dagger}_j \rangle  - 2\langle L_i \rangle \langle L^{\dagger}_i \rangle \langle L_j \rangle \langle L^{\dagger}_j \rangle + \langle L_i L_j \rangle \langle L^{\dagger}_i L^{\dagger}_j \rangle - 1  \right)
\end{equation}
where $\langle \cdot \rangle$ is taken with respect to $|\psi\rangle_{QR}$, that is, $\langle \cdot \rangle = \frac{1}{D}\mathrm{Tr}(\cdot P)$.

Now we turn to $\rho_{Q}(p)$, which takes the same form as Eq.~\eqref{eq:expansion_2nd_order} but with $QR$ substituted as $Q$. The initial state is $\rho_{Q}(0)= \frac{1}{D}P$ where $P$ is the projector onto the code subspace. Thus $\rho_{Q}(0)^2 = \frac{1}{D} \rho_{Q}(0)$. The first $D$ eigenvalues of $\rho_{Q}(p)$ can be expanded as $1/D-p\alpha_{k} + p^2 \alpha^{(2)}_{k}$, where $k=1,2,\cdots,D$. The entropy $S_Q$ can be expanded as 
\begin{equation}
    S_Q = \log D + \left(\sum_{k=1}^D \alpha_k\right) p\log p - \left(\sum_{k=1}^D \alpha^{(2)}_k\right) p^2\log p + \mathrm{regular}
\end{equation}
We can use second order perturbation theory to compute the coefficient in front of $p^2\log p$. 
\begin{equation}
\label{eq:SQ2nd_1}
    \sum_{k=1}^D \alpha^{(2)}_k = D\mathrm{Tr}(P \delta P (1-P) \delta P) + \frac{1}{2}\mathrm{Tr}(P \delta^2 P),
\end{equation}
where the factor of $D$ comes from the eigenvalue $1/D$ in the unperturbed $\rho_Q$ and we have denoted
\begin{equation}
\label{eq:SQ2nd_2}
       \delta P = \frac{1}{D}\sum_{i} \frac{L_i P L^{\dagger}_i-P}{2}, ~~ \delta^2 P = \frac{1}{D}\sum_{i,j} \frac{L_i L_j P L^{\dagger}_j L^{\dagger}_i - P}{4}.
\end{equation}
Substituting Eq.~\eqref{eq:SQ2nd_2} into Eq.~\eqref{eq:SQ2nd_1}, we obtain
\begin{equation}
    \sum_{k=1}^D \alpha^{(2)}_k = \frac{1}{8D}\sum_{i,j} \left( 2\mathrm{Tr}(PL_i P L^{\dagger}_i L_j P L^{\dagger}_j)  - 2\mathrm{Tr}(P L_i P L^{\dagger}_i P L_j P L^{\dagger}_j) + \mathrm{Tr}( P L_i L_j P L^{\dagger}_i L^{\dagger}_j) - D  \right)
\end{equation}
Now we can identify the coefficient $b^{(2)}$ in Eq.~\eqref{eq:Ic_expansion_2nd_order} as
\begin{equation}
    b^{(2)} = \sum_{k=1}^D \alpha^{(2)}_k -\nu^{(2)}_0  := b^{(2)}_1 + b^{(2)}_2 + b^{(2)}_3,
\end{equation}
where we separate the sum into three terms,
\begin{eqnarray}
\label{eq:2nd_perb_1}
    b^{(2)}_1 &=& \frac{1}{4D^4}\sum_{i,j} \left( D^3\mathrm{Tr}(PL_i P L^{\dagger}_i P L_j P L^{\dagger}_j)-\mathrm{Tr} (P L_i)  \mathrm{Tr}(P L^{\dagger}_i)\mathrm{Tr} (P L_j) \mathrm{Tr}(P L^{\dagger}_j) \right) \\
    \label{eq:2nd_perb_2}
    b^{(2)}_2 &=& \frac{1}{4D^3}\sum_{i,j} \left( D^2\mathrm{Tr}(PL_i P L^{\dagger}_i L_j P L^{\dagger}_j)- \mathrm{Tr} (P L_i)  \mathrm{Tr}(P L^{\dagger}_i L_j) \mathrm{Tr}(P L^{\dagger}_j) \right)\\
    \label{eq:2nd_perb_3}
    b^{(2)}_3 &=& \frac{1}{8D^2} \sum_{i,j} \left( D\mathrm{Tr}( P L_i L_j P L^{\dagger}_i L^{\dagger}_j) - \mathrm{Tr} (P L_i L_j) \mathrm{Tr} (P L^{\dagger}_i L^{\dagger}_j) \right)
\end{eqnarray}
Again the result makes sense since $b^{(2)}=0$ if and only if the Knill-Laflamme condition is satisfied to the second order, $PL^{\dagger}_i L_j P \propto P$ and $PL_i L_j P \propto P$. The first condition says that any codeword state with one error can be distinguished from any other codeword state with one error. The second condition says that any codeword state with two errors is distinguishable from any other codeword states with no error. 

Now we analyze the scaling of each term for the CFT code. For the first term $b^{(2)}_1$, it scales as $O(n^{2-4\Delta})$. Combining with $p^2$ gives $(pn^{1-2\Delta})^2$, thus this term does not change the exponent from the first-order perturbation. For the second term $b^{(2)}_2$, we can separate the sum into $i=j$ and $i\neq j$ pieces. The $i=j$ piece is identical to Eq.~\eqref{eq:b_compact} up to a constant factor. 
For the $i\neq j$ piece, we first analyze the scaling of the term
\begin{equation}
   \sum_{i\neq j} \mathrm{Tr}(P L^{\dagger}_i L_j) = n \sum_{\alpha=1}^D \sum_{i=1}^{n-1} \langle \phi_{\alpha} | L^{\dagger}_0 L_i|\phi_{\alpha}\rangle,
\end{equation}
where we have made use of translation invariance. Given a lattice realization of the CFT, we can do OPE on the lattice with $O^{\dagger}(i) O(j) \sim |i-j|^{-2\Delta+\Delta_{O^{\dagger}O}} :O^{\dagger}O:((i+j)/2) + \cdots$, where we recall that $\Delta$ denotes the scaling dimension of $O$. Now the sum over $i$ can be estimated based on whether  $\Delta_{O^{\dagger} O}-2\Delta>-1$. If $\Delta_{O^{\dagger} O}-2\Delta>-1$, then the sum diverges as $n\rightarrow \infty$, then 
\begin{equation}
\sum_{i\neq j} \mathrm{Tr}(P L^{\dagger}_i L_j) \propto n^{2-2\Delta} ~~ (\Delta_{O^{\dagger}O}<2\Delta-1)
\end{equation}
otherwise we have a converge sum in $n$ which gives rise to
\begin{equation}
    \sum_{i\neq j} \mathrm{Tr}(P L^{\dagger}_i L_j) \propto n^{1-\Delta_{O^{\dagger}O}} ~~ (\Delta_{O^{\dagger}O}>2\Delta-1)
\end{equation}
In either case, we consistently have
\begin{equation}
    \sum_{i\neq j} \mathrm{Tr}(P L^{\dagger}_i L_j)\leq O( n^{2-2\Delta})
\end{equation}
Taking into account the other factor $\mathrm{Tr}(PL_i) \propto n^{-\Delta}$, we conclude that the second term $b^{(2)}_2 \leq O(n^{2-4\Delta})$ cannot give a larger exponent than the first-order perturbation. The first term in $b^{(2)}_2$ can be argued analogously.

Finally, we consider $b^{(2)}_3$. Following a similar derivation as above, we use the OPE $O(i) O(j) \sim |i-j|^{-2\Delta+\Delta^{(2)}} :O^2:((i+j)/2) + \cdots$. We can obtain
\begin{equation}
b^{(2)}_3 \propto
    \begin{cases} 
n^{2-4\Delta} & \text{if } 2\Delta^{(2)}>1+4\Delta \\
n^{1-2\Delta^{(2)}} & \text{if } 2\Delta^{(2)}<1+4\Delta
\end{cases}
\end{equation}
If $\Delta^{(2)}<\Delta$, then clearly the second condition is satisfied and $b^{(2)}_3 \propto n^{1-2\Delta^{(2)}}$. This term has larger exponent compared to the first-order perturbation scaling $n^{1-2\Delta}$. Thus if $n$ is large we obtain $\delta I_c \propto p^2 n^{1-2\Delta^{(2)}} \log p$ and $\nu = (1-2\Delta^{(2)})/2$.

\subsection{Higher order perturbation}
Our main claim is that the noise is correctable at a finite threshold if and only if $\Delta_{\min}>1/2$, where $\Delta_{\min}$ is the minimum of nonzero scaling dimensions in the fusion algebra of the jump operators.  So far we have shown it within first and second order perturbation theory. We have argued for the ``only if" direction to all orders using the RG argument in the last section. For the ``if" direction one has to work out the full perturbation theory to all orders, but it will be technically very difficult. Instead, based on the second-order perturbation result, we give an argument in this section that $\Delta_{\min}>1/2$ implies $\nu<0$ to all orders, which in turn implies that the noise is correctable.

The singular term in the coherent information can be expanded in all orders in $p$,
\begin{equation}
    I_c = \log D +\sum_{r=1}^{\infty} b^{(r)} p^r \log p + \mathrm{regular}.
\end{equation}
Each of the coefficient $b^{(r)}\propto n^{1-2\Delta^{(r)}}$. The exponent $\nu$ is determined by the smallest $\Delta^{(r)}$ in the expansion by $\nu = (1-2\Delta^{(r)})/r$.  We will argue that $\Delta^{(r)}$ is the lowest scaling dimension of the first $r$-th order OPE of the jump operator below.

As we have shown, to the first-order $r=1$ and second order $r=2$, $b^{(r)}$ reproduces part of the Knill-Laflamme conditions, see Eq.~\eqref{eq:b_compact} and Eqs.~\eqref{eq:2nd_perb_1}-\eqref{eq:2nd_perb_3}. This feature should persist to all orders, because the Knill-Laflamme condition is equivalent to $I_c = \log D$, which follows from the fact that both are equivalent to exact decodability. Furthermore, to the $r$-th order, there are $2r$ jump operator insertions between the $P$ operator inside the trace. This observation allows us to write down the form of $b^{(r)}$ without an explicit calculation. Take $r=3$ as an example, the part of Knill Laflamme conditions are
\begin{equation}
\label{eq:KL_3rd_order}
    PL_i L_j L_k P \propto P, ~PL^{\dagger}_i L_j L_k P \propto P,~PL^{\dagger}_i L^{\dagger}_j L_k P \propto P, ~PL^{\dagger}_i L^{\dagger}_j L^{\dagger}_k P \propto P
\end{equation}
One can then write down the following terms which could contribute to $b^{(3)}$,
\begin{eqnarray}
    b^{(3)}_1 \propto &\sum_{i,j,k}& [D\tr(PL_i L_j L_k P L^{\dagger}_i L^{\dagger}_j L^{\dagger}_k) - \tr(PL_i L_j L_k) \tr(P L^{\dagger}_i L^{\dagger}_j L^{\dagger}_k)] \nonumber \\
   b^{(3)}_2 \propto &\sum_{i,j,k}& [D^2\tr(PL_iPL^{\dagger}_i L_j L_k P L^{\dagger}_j L^{\dagger}_k) - \tr(PL_i ) \tr(P L^{\dagger}_i L_j L_k ) \tr(P L^{\dagger}_j L^{\dagger}_k)] + h.c.\nonumber \\
   b^{(3)}_3\propto &\sum_{i,j,k}& [D^2\tr(PL^{\dagger}_i L_jPL^{\dagger}_j L_k P L^{\dagger}_k  L_i) - \tr(PL^{\dagger}_i L_j) \tr(P L^{\dagger}_j L_k ) \tr(P L^{\dagger}_k L_i)] \nonumber \\
   b^{(3)}_4 \propto&\sum_{i,j,k}& [D^3\tr(PL^{\dagger}_i P L_jPL^{\dagger}_j L_k P L^{\dagger}_k  L_i) - \tr(PL^{\dagger}_i) \tr( PL_j) \tr(P L^{\dagger}_j L_k ) \tr(P L^{\dagger}_k L_i)] \nonumber \\
   b^{(3)}_5\propto &\sum_{i,j,k}& [D^4\tr(PL^{\dagger}_i P L_jPL^{\dagger}_j P L_k P L^{\dagger}_k  L_i) - \tr(PL^{\dagger}_i) \tr( PL_j) \tr(P L^{\dagger}_j) \tr (P L_k ) \tr(P L^{\dagger}_k L_i)] \nonumber \\
   b^{(3)}_6 \propto&\sum_{i,j,k}& [D^5\tr(PL^{\dagger}_i P L_jPL^{\dagger}_j P L_k P L^{\dagger}_k P L_i) - \tr(PL^{\dagger}_i) \tr( PL_j) \tr(P L^{\dagger}_j) \tr (P L_k ) \tr(P L^{\dagger}_k) \tr(P L_i)] \nonumber 
\end{eqnarray}
Denoting $\Delta^{(3)}_{\min} = \min\{\Delta,\Delta^{(2)},\Delta^{(3)}\}$ and assuming $\Delta^{(3)}_{\min}>1/2$, we can argue that each term above is bounded above by $b^{(3)}_i \leq O(n^{1-2\Delta^{(3)}_{\min}})$. In fact, if $\Delta^{(3)}$ is the smallest, only the first term $b^{(3)}_1$ saturates the bound. The rest of terms are controlled by $\Delta$ and $\Delta^{(2)}$ as in previous orders of perturbation theory. For example one can show that $b^{(3)}_6 \propto n^{3-6\Delta} \leq O(n^{1-2\Delta^{(3)}_{\min}})$ and $b^{(3)}_4 \leq O( n^{1-2\Delta}) \leq O( n^{1-2\Delta^{(3)}_{\min}})$. Now we turn to $b^{(3)}_1$. We will argue that the largest contribution comes from when $i,j,k$ are all close to each other, where we can fuse $L_i,L_j,L_k$ to one error with scaling dimension $\Delta^{(3)}$ and thus gives $O(n^{1-2\Delta^{(3)}}) = O(n^{1-2\Delta^{(3)}_{\min}})$. If the three points are $O(n)$ far away, then their fusion gives rise to a factor of $n^{-3\Delta}$. Each points can take $O(n)$ locations, so we have in total $n^3 \times (n^{-3\Delta})^2 = n^{3-6\Delta}$ contributions to $b^{(3)}_1$. Likewise, if two points are close and the other point is $O(n)$ far away, the contribution to $b^{(3)}_1$ would be $n^2 \times (n^{-\Delta^{(2)}-\Delta})^2 = n^{2-2\Delta^{(2)}-2\Delta} < n^{2-4\Delta^{(3)}_{\min}}<n^{1-2\Delta^{(3)}_{\min}}$.

Now we can generalize the above argument to arbitrary orders. Let $\Delta_{\min}:=\min_{r}\{\Delta^{(r)}\}$ be the minimal scaling dimension and the minimum is achieved in the $r$-th order (if there are multiple orders which achieve the minimum, then we choose $r$ to be the lowest order).
The leading contribution to $b^{(r)}$ is given by
\begin{equation}
\label{eq:r-perturb}
    b^{(r)}\propto \sum_{i_1,i_2,\cdots i_r} D\tr(PL_{i_1}L_{i_2}\cdots L_{i_r}PL^{\dagger}_{i_1}L^{\dagger}_{i_2}\cdots L^{\dagger}_{i_r}) - \tr(PL_{i_1}L_{i_2}\cdots L_{i_r})\tr(PL^{\dagger}_{i_1}L^{\dagger}_{i_2}\cdots L^{\dagger}_{i_r}).
\end{equation}
Following a similar argument as $b^{(3)}_1$ above, if $\Delta^{(r)}$ achieves the minimum, then dominant contribution comes from when all $i$'s are close to each other. This the gives $b^{(r)}\propto n^{1-2\Delta^{(r)}}$ and consequently $\nu = (1-2\Delta^{(r)})/r$. 


\section*{Proving at least $O(\log\log n)$ logical qubits}
The main result that we will prove is that
\begin{equation}
    b(n) = \frac{1}{D^2}\sum_{i=1}^n[D\tr(L_i P L^{\dagger}_i P)-\tr(L_i P)\tr(L^{\dagger}_i P)] \rightarrow 0
\end{equation}
if the jump operator $L_i$ corresponds to a CFT primary operator $O$ with scaling dimension $\Delta_0>1/2$ and
\begin{equation}
    P = \sum_{m,\bar{m}=0}^{M} |\partial^m \bar{\partial}^{\bar{m}} \phi\rangle\langle\partial^m \bar{\partial}^{\bar{m}} \phi|
\end{equation}
for $M= \mathrm{polylog}(n)$. Recall that the condition $b(n)\rightarrow 0$ implies that the code can correct the dephasing noise at a finite threshold (assuming the scaling hypothesis). With this choice of code subspace we have $D=M^2$. Thus, this ensures at least $k=O(\log \log n)$ logical qubits. In order for proof, one first observes that
\begin{equation}
    b(n) \leq \frac{1}{D} \tr(L_i P L^{\dagger}_i P)
\end{equation}
so we only need to prove
\begin{equation}
    \sum_{i=1}^n\tr(L_i P L^{\dagger}_i P) < O(M^2)
\end{equation}
The matrix element of $L_i$ in the code subspace must take the following form 
\begin{equation}
    |\langle \partial^p \bar{\partial}^{\bar{p}} \phi|L_i |\partial^q\bar{\partial}^{\bar{q}} \phi\rangle| = \left(\frac{2\pi}{n}\right)^{\Delta_0}C_{O\phi\phi}f(p,q)f(\bar{p},\bar{q})
\end{equation}
due to conformal symmetry. The dependence on $i$ is eliminated because of translation symmetry. Thus, we only need to prove
\begin{equation}
\label{eq:codespace_cond}
    \sum_{p,q=0}^{M} f(p,q)^2 = o(n^{2\Delta_0-1}).
\end{equation}

Below we compute $f(p,q)$ using conformal field theory techniques. First, let us write the overlap in terms of correlation functions by the Weyl transformation,
\begin{equation}
    \langle \partial^p \bar{\partial}^{\bar{p}} \phi|L_i |\partial^q\bar{\partial}^{\bar{q}} \phi\rangle = \left(\frac{2\pi}{n}\right)^{\Delta_0} \frac{\langle (\partial^p \bar{\partial}^{\bar{p}} \phi(0))^{\dagger} O(z_1=1) \partial^q \bar{\partial}^{\bar{q}} \phi(0) \rangle}{\sqrt{N_{p,\bar{p}} N_{q,\bar{q}} }},
\end{equation}
where
\begin{equation}
    N_{p,\bar{p}} = \langle (\partial^p \bar{\partial}^{\bar{p}} \phi(0))^{\dagger} \partial^p \bar{\partial}^{\bar{p}} \phi(0)\rangle
\end{equation}
is a normalization constant. We will compute the denominator first, which goes as below
\begin{eqnarray}
    N_{p,\bar{p}} &=& \lim_{z\rightarrow \infty,z_0\rightarrow 0} \langle (-z^2\partial)^p (-\bar{z}^2\bar{\partial})^{\bar{p}} (z^{2h}\bar{z}^{2\bar{h}}\phi(z,\bar{z}))\partial^p_0 \bar{\partial}^{\bar{p}}_0 \phi(z_0,\bar{z}_0)\rangle \\
    &=& \lim_{z\rightarrow \infty,z_0\rightarrow 0} (-z^2\partial)^p (-\bar{z}^2\bar{\partial})^{\bar{p}} \partial^p_0 \bar{\partial}^{\bar{p}}_0 \langle  z^{2h}\bar{z}^{2\bar{h}}\phi(z,\bar{z}) \phi(z_0,\bar{z}_0)\rangle \\
    &=& \lim_{z\rightarrow \infty,z_0\rightarrow 0} [(-z^2\partial)^p   (z^{2h} \partial^p_0 (z-z_0)^{-2h})]\times c.c. \\
    &=& \lim_{z\rightarrow \infty} \frac{\Gamma(p+2h)}{\Gamma(2h)}[(-z^2\partial)^p   (z^{2h}  z^{-2h-p})]\times c.c. \\
     &=& \lim_{z\rightarrow \infty} \frac{\Gamma(p+2h)}{\Gamma(2h)}[(-z^2\partial)^p  z^{-p} ]\times c.c. \\
     &=& \lim_{w\rightarrow 0} \frac{\Gamma(p+2h)}{\Gamma(2h)}[\partial^p_w  w^{p} ]\times c.c. \\
     &=& p!\frac{\Gamma(p+2h)}{\Gamma(2h)} \times c.c.
\end{eqnarray}
In this calculation we have used the fact that $\phi(0)^{\dagger} = \lim_{z\rightarrow\infty} z^{2h} \bar{z}^{2\bar{h}} \phi(z,\bar{z})$ and in the last step we have substitute $w=1/z$ in the limit. Note that here $c.c.$ means substitute $p$ with $\bar{p}$ and $h$ with $\bar{h}$.

In a similar fashion, we can compute the numerator as
\begin{eqnarray}
    &~&\langle (\partial^p \bar{\partial}^{\bar{p}} \phi(0))^{\dagger} O(z_1=1) \partial^q \bar{\partial}^{\bar{q}} \phi(0) \rangle \\
    &=& \lim_{z\rightarrow \infty,z_0\rightarrow 0} (-z^2\partial)^p (-\bar{z}^2\bar{\partial})^{\bar{p}} \partial^q_0 \bar{\partial}^{\bar{q}}_0 \langle z^{2h} \bar{z}^{2\bar{h}} \phi(z,\bar{z}) O(z_1 = 1) \phi(z_0,\bar{z}_0)\rangle \\
    &=& C_{O\phi\phi}\lim_{z\rightarrow \infty,z_0\rightarrow 0}(-z^2\partial)^p [z^{2h} \partial^q_0 ((z-z_0)^{h_0-2h} (z-1)^{-h_0} (1-z_0)^{-h_0})] \times c.c. \\
    &=& C_{O\phi\phi} \left[\sum_{k=0}^{\min(p,q)} \frac{q!}{(q-k)!} \frac{\Gamma(h_0+q-k)\Gamma(2h-h_0+k) \Gamma(h_0+p-k)}{\Gamma(h_0)^2\Gamma(2h-h_0)}\right]\times c.c.
\end{eqnarray}
We have omitted the details of evaluating the limit and derivatives. Now we can identify the expression of $f(p,q)$,
\begin{equation}
    f(p,q) = \frac{\Gamma(2h)}{\sqrt{p! q! \Gamma(p+2h)\Gamma(q+2h)}} \sum_{k=0}^{\min(p,q)} \frac{q!}{(q-k)!} \frac{\Gamma(h_0+q-k)\Gamma(2h-h_0+k) \Gamma(h_0+p-k)}{\Gamma(h_0)^2\Gamma(2h-h_0)}.
\end{equation}
We now show that, if $h,h_0$ are $O(1)$ numbers, then
\begin{equation}
    f(p,q) \leq \mathrm{poly}(p,q)
\end{equation}
We use the fact that $ (n+\lfloor x \rfloor-1)!\leq \Gamma(n+x) \leq (n+\lfloor x \rfloor)!$ for integer $n\geq 2$ and positive real number $x$, we can estimate that $\Gamma(h_0+q-k) = \mathrm{poly}(q)\times (q-k)!$. In a similar way, we make the substitution $\Gamma(2h-h_0+k) =  \mathrm{poly}(k)\times k!$. Thus we estimate the term in the sum as 
\begin{equation}
    \frac{q!}{(q-k)!} \frac{\Gamma(h_0+q-k)\Gamma(2h-h_0+k) \Gamma(h_0+p-k)}{\Gamma(h_0)^2\Gamma(2h-h_0)} = \mathrm{poly}(p,q) \times q! k! (p-k)!
\end{equation}
The sum in $k$ can be upper and lower bounded because $\sum_{k=0}^p k! (p-k)! = O(1) \times p!$, thus we obtain $f(p,q) \leq \mathrm{poly}(p,q)$. As long as $M=\mathrm{polylog}(n)$, the LHS Eq.~\eqref{eq:codespace_cond} scales smaller than any power law. This completes the proof.

\section*{Free fermion CFT code}
We further justify our $\nu = 1-2\Delta$ prediction with an example where the coherent information can be computed exactly. We consider the free Majorana fermion CFT realized by
\begin{equation}
    H = i\sum_{i=1}^{2n} c_i c_{i+1}, 
\end{equation}
where $\{c_i,c_j\} = 2\delta_{ij}$ represents the Majorana fermions. This model is dual to the TFIM under the Jordan-Wigner transformation. We take the Neveu-Schwarz boundary condition $c_{2i+1} = -c_{1}$. The code subspace is taken to be spanned by the ground state (corresponding to the identity operator $\mathbb{I}$) and the first excited state (corresponding to the chiral fermion operator $\Psi$). Note that the two codewords have different fermionic parity.

\begin{figure}
    \centering
    \includegraphics[width = 0.6\linewidth]{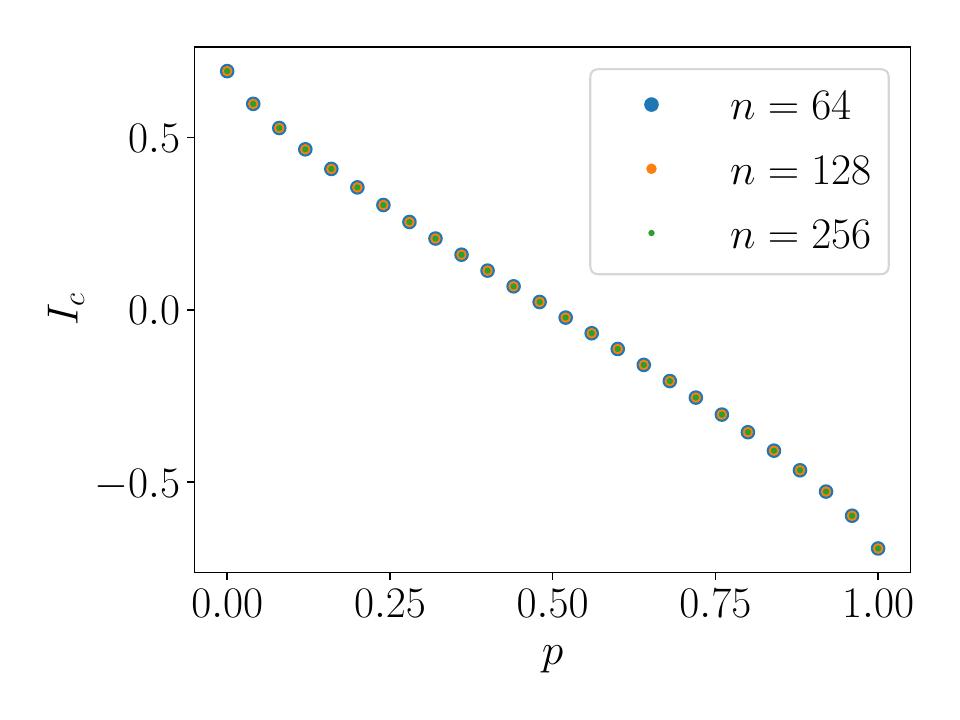}
    \caption{Coherent information for the free fermion CFT code under amplitude damping noise}
    \label{fig:FF}
\end{figure}

We consider the amplitude damping channel as in Ref.~\cite{zou2023channeling}. The Lindbladian
\begin{equation}
        \mathcal{L}(\rho) = \sum_{i=1}^n \left[L_i \rho L^{\dagger}_i-\frac{1}{2}\{L^{\dagger}_i L_i,\rho\}\right]
\end{equation}
has local jump operators 
\begin{equation}
    L_j = a_j,
\end{equation}
where $a_j = (c_{2j-1} + i c_{2j})/2$ is the complex fermion annihilation operator. Now we can integrate the Lindbladian to give a channel $\mathcal{N}_p = e^{t \mathcal{L}}$, where $p=1-e^{-t}$ is the damping probability.
In terms of Kraus representation, the single-site channel can be written as
\begin{equation}
    \mathcal{N}^{[j]}_p(\rho) = K_0 \rho K^{\dagger}_0 + K_1 \rho K^{\dagger}_1
\end{equation}
where $K_1 = \sqrt{p} a$ and $K_0 = I + (\sqrt{1-p}-1)a^{\dagger} a $.
The jump operator has scaling dimension $\Delta_{a} =1/2$ and thus one expect that $\nu = 1-2\Delta_{a} = 0$. Thus, the coherent information only depends on $p$ and does not depend on $n$. It turns out that here the coherent information can be computed using correlation matrix techniques, since $|\psi_{RQ}\rangle$ is a Gaussian state and $\mathcal{N}^{[j]}_p$ is a Gaussian channel. We indeed observe that the $I_c$ only depends on $p$, see Fig.~\ref{fig:FF}.

This means that the fermionic amplitude damping noise is not correctable for $p\neq 0$ in the thermodynamic limit, as $I_c < \log D$. However, as opposed to the case where $\Delta<1/2$, the coherent information is a continuous function of $p$ and does not jump immediately at $p=0$. Thus, for small $p$ the noise is still approximately correctable for any size $n$.  
\section*{Renyi coherent information for CFT code}
In this section we consider the Renyi coherent information, defined by
\begin{equation}
    I^{(n)}_c = S^{(n)}_Q - S^{(n)}_{QR},
\end{equation}
where
\begin{equation}
    S^{(n)}(\rho):= \frac{1}{1-n}\log \tr(\rho^n),
\end{equation}
We show that the Renyi coherent information still has a scaling collapse $I^{(n)}_c = f(p N^{\nu^{(n)}})$, where $\nu^{(n)}$ could in general be different across different $n$. Note that in the section we will use $N$ to denote the total number of physical qubits and $n$ to denote the Renyi index. We show that for each Renyi index $n$ there is a notion of relevant, irrelevant and marginal dephasing based on defect RG flow in the CFT. This defect RG flow is identical to what we encounter in Ref.~\cite{zou2023channeling}. We will first review the result in Ref.~\cite{zou2023channeling} and show that the RG flow of coherent information is identical to the defect RG flow. We will present the numerical result on the Ising model and show that the $Z$ dephasing is marginal at $n=2$, which means $\nu^{(2)}=0$ and thus $I^{(2)}_c$ is independent of $n$. Recall that we have previously shown that $Z$ dephasing is correctable with $\nu = -1$ in the replica limit $n\rightarrow 1$. Thus, this section also serves as an example where the Renyi index $n=2$ gives a qualitatively different prediction from the replica limit. 
\subsection{Review of channeling quantum criticality}
We consider the entropy of a subsystem $A$ of a CFT eigenstate $|\phi_{\alpha}\rangle$ under decoherence channel $\mathcal{N}$. Here we have generalized the approach in Ref.~\cite{zou2023channeling} which only considers the CFT ground state. The mixed state of interest is $\rho_{\alpha} = \mathcal{N}(|\phi_{\alpha}\rangle \langle \phi_{\alpha}|)$. The $n-$th momentum of the reduced density matrix is
\begin{eqnarray}
    \tr(\rho^n_{\alpha,A}) &=& \tr(\rho_{\alpha}^{\otimes n} \tau^{(n)}_{A} \bo_B) \\
    &=& \tr((|\phi_{\alpha}\rangle \langle \phi_{\alpha}|)^{\otimes n} \mathcal{N}^{*\otimes n}(\tau^{(n)}_{A})\bo_B) \\
    &=& \langle \phi^{\otimes n}_{\alpha}|\mathcal{B}^{(n)}_{\mathcal{N},A} \bo_B|\phi^{\otimes n}_{\alpha}\rangle \\
    &=& \langle \phi^{\otimes 2n}_{\alpha}|\mathcal{B}^{(n)}_{\mathcal{N},A} \bo_B\rangle
\end{eqnarray}
where $B$ is the complement of $A$, $\tau^{(n)}$ is the forward permutation on replicas and $\mathcal{B}^{(n)}_{\mathcal{N}} = \mathcal{N}^{*\otimes n}(\tau^{(n)}) $. In the last step we have made use of the duality between operators and states in doubled Hilbert space and the overlap is taken within $2n$ copies of CFT. In terms of the path integral, this can be represented by a three-point correlation function of a bulk operator and two boundary condition changing operators. Under RG flow, the boundary state $|\mathcal{B}^{(n)}_{\mathcal{N}}\rangle$ flows to the same conformal boundary condition regardless of the bulk insertion. This means that the RG flow of the channel defined in Ref.~\cite{zou2023channeling} can be applied to all low-energy eigenstates. In particular, if the channel $\mathcal{N}$ flows to complete dephasing, then the Renyi negativity of all low-energy states becomes area law under the channel. If the channel is irrelevant (i.e., flows to the identity channel), then the entanglement scaling of all low-energy eigenstates at long-distances remains unchanged (except non-universal constants).  

\subsection{RG flow of the Renyi coherent information}
So far we have shown that the RG flow of quantum channels are identical for all low-energy states.  One may naturally expect that the code property of the low-energy subspace subject to uniform decoherence follows the same RG flow. If the channel flows to complete dephasing, then the error cannot be corrected. If the channel flows to the identity channel, then the error can be corrected. The two cases correspond to relevant and irrelevant dephasing, respectively. The correctability of the error is characterized by the loss of coherent information. The subtlety is that Ref.~\cite{zou2023channeling} works at integer Renyi index $n\geq 2$ but the code property is only related to the replica limit $n\rightarrow 1$. 

We will fix a Renyi index $n\geq 2$ below. We compute the Renyi coherent information of the CFT code and show that it decreases to zero for relevant dephasing and remains invariant for irrelevant dephasing. For marginal case, the Renyi coherent information does not depend on system size $N$ and is a continuous function of $p$.

we first couple the CFT to the reference qudit,
\begin{equation}
    |\psi\rangle_{QR} = \frac{1}{\sqrt{D}}\sum_{\alpha=1}^D|\phi_{\alpha}\rangle_Q \otimes |\alpha\rangle_R 
\end{equation}
The state subjected to decoherence is $\rho_{QR} = \mathcal{N}_Q(|\psi\rangle\langle \psi|)$. The Renyi coherent information is
\begin{eqnarray}
\label{eq:Ic}
    I^{(n)}_c = \frac{1}{n-1}\log \frac{\tr\rho^n_{QR}}{\tr\rho^n_Q} 
\end{eqnarray}
In terms of the channeled twist operator $\mathcal{B}^{(n)}_{\mN}$, the numerator and demominator can be expressed as
\begin{eqnarray}
\label{eq:SQR}
    \tr\rho^n_{QR} &=& \frac{1}{D^n}\sum_{\alpha_1,\cdots \alpha_n=1}^D  \left\langle\bigotimes_i \phi_{\alpha_{\tau(i)}}\right|\mathcal{B}^{(n)}_{\mN} \left|\bigotimes_{i}\phi_{\alpha_{i}}\right\rangle. \\
\label{eq:SQ}
    \tr\rho^n_{Q} &=& \frac{1}{D^n}\sum_{\alpha_1,\cdots \alpha_n=1}^D\left\langle \bigotimes_{i}\phi_{\alpha_i}\right|\mathcal{B}^{(n)}_{\mN}\left|\bigotimes_{i}\phi_{\alpha_i}\right\rangle,
\end{eqnarray}
where $\tau(i) = i+1 ~\mathrm{mod}~ n$ forward permutes the replica index. If the channel flows to identity channel, then $\mathcal{B}^{(n)}_{\mN}$ flows to the twist operator $\tau^{(n)}$. In this case the Eq.~\eqref{eq:SQR} gets contributions from all terms and Eq.~\eqref{eq:SQ} only gets contributions from the diagonal terms (where all $\alpha_i$'s are equal). Thus, $\tr\rho^n_{QR}=1$ and $\tr\rho^n_{Q} = D^{1-n}$, and
\begin{equation}
\label{eq:caseIc1}
    I^{(n)}_c = \log D, ~~~\mathrm{irrelevant ~dephasing.}
\end{equation}
If instead the channel flows to complete dephasing, say, in the direction of $\sigma_{\vec{n}} \equiv \vec{n}\cdot \vec{\sigma}$, then the on each site
\begin{equation}
    \mathcal{B}^{(n)}_{\mN} = \sum_{i=+,-} |i^{\otimes n}\rangle\langle i^{\otimes n}|, ~~~\mathrm{complete ~dephasing.}
\end{equation}
where $|\pm\rangle$ is the eigenvector of $\sigma_{\vec{n}}$ with eigenvalue $\pm 1$.  Since $\mathcal{B}^{(n)}_{\mN}$ is a completely symmetric tensor with respect to permutation of the $2n$ replicas, each term in Eq.~\eqref{eq:SQR} and Eq.~\eqref{eq:SQ} is equal. Thus, the Renyi coherent information Eq.~\eqref{eq:Ic} vanishes for all $n\geq 2$, i.e.,
\begin{equation}
\label{eq:caseIc2}
    I^{(n)}_c = 0, ~~~\mathrm{complete ~dephasing.}
\end{equation}
Finally, the extreme case where we have complete depolarization (erasure) gives $\mathcal{B}_{\mathcal{N}} = \bo$. Thus, only diagonal terms contribute to Eq.~\eqref{eq:SQR} and
\begin{equation}
\label{eq:caseIc3}
    I^{(n)}_c = -\log D, ~~~\mathrm{complete ~erasure.}
\end{equation}

Thus we have established that for a given Renyi index $n$, if the dephasing channel is relevant, then $I^{(n)}_c = 0$ in the thermodynamic limit; if the dephasing channel is irrelevant, then $I^{(n)}_c = \log D$ in the thermodynamic limit. 
\subsection{Numerical result for the Ising model}

\begin{figure}[htbp]
    \centering
    \includegraphics[width=0.3\linewidth]{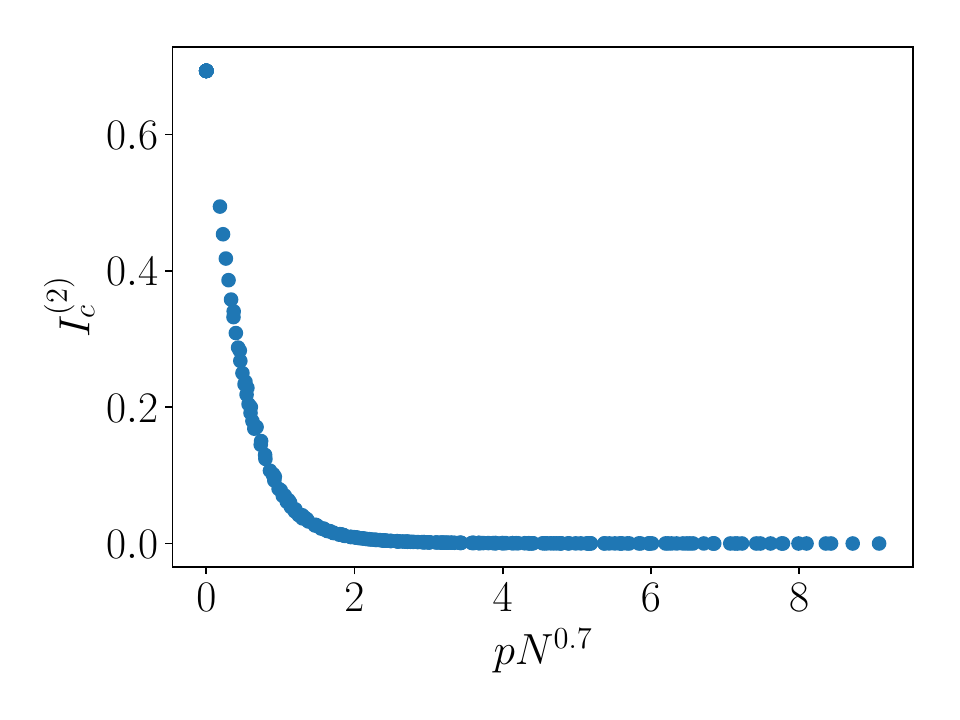}
    \includegraphics[width=0.3\linewidth]{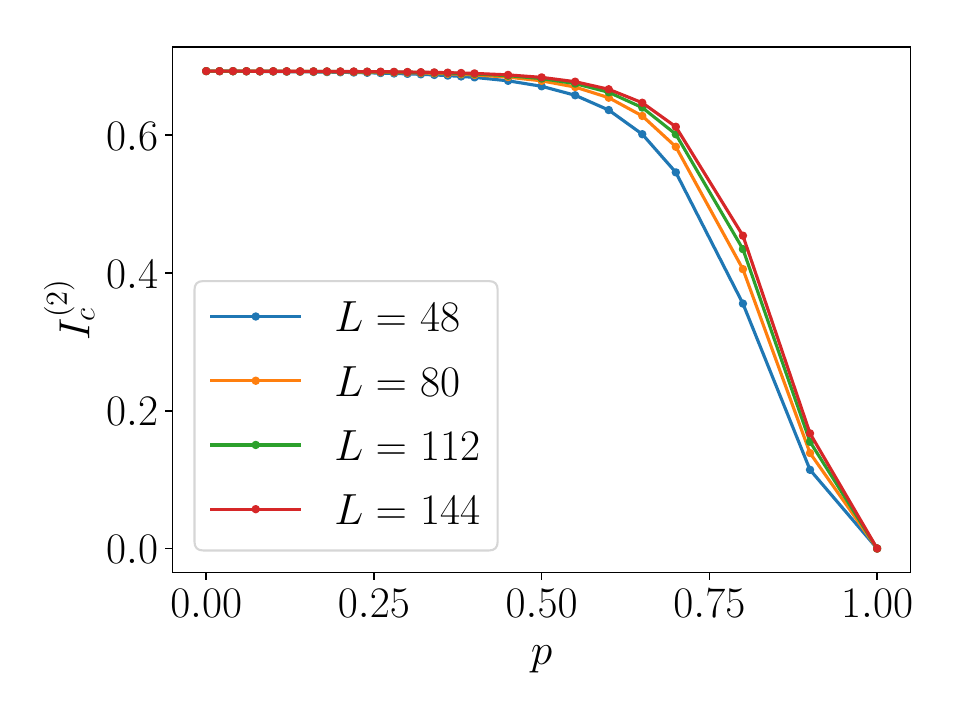}
    \includegraphics[width=0.3\linewidth]{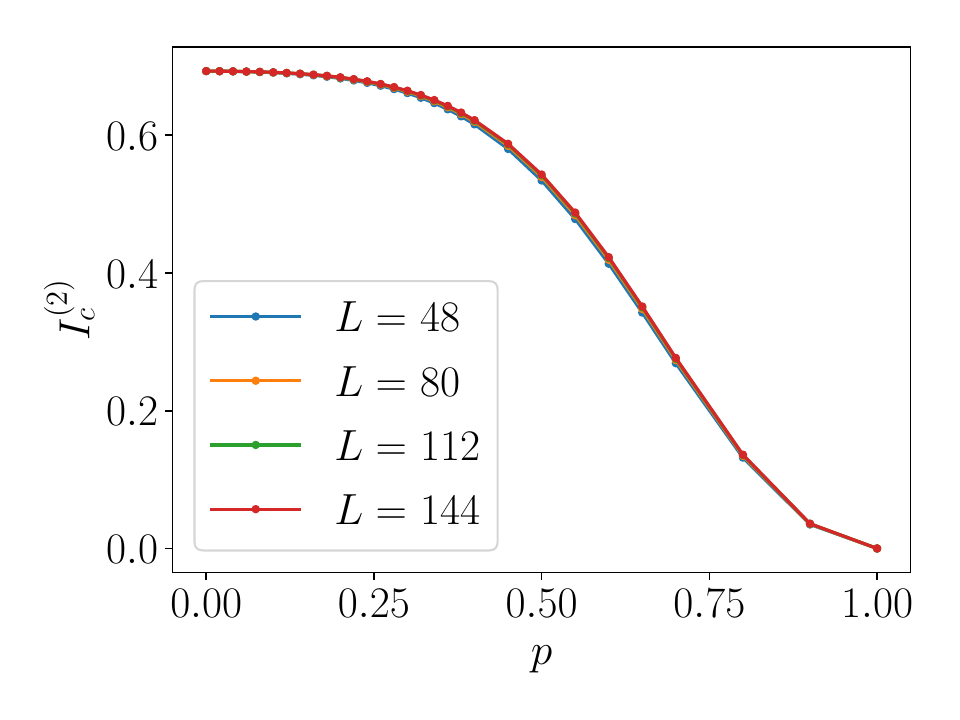}
    \caption{Renyi coherent information of the Ising CFT code under $X$ (left), $Y$ (center) and $Z$ (right) dephasing}
    \label{fig:CI2}
\end{figure}

Taking the $|I\rangle$ and $|\sigma\rangle$ as the codeword states, we consider the Renyi-2 coherent information under $X$,$Y$ or $Z$ dephasing for the Ising CFT code. We obtain the following results (see Fig.~\ref{fig:CI2}).
The result indicates that the $X,Y$ and $Z$ dephasings are relevant, irrelevant and marginal respectively, which can be seen in the following way. For $X$ dephasing, increasing the system size decreases the coherent information. For $Y$ dephasing, increasing the system size increases $I^{(2)}_c$ to $\log 2$ at small $p$. The $Z$ dephasing is the marginal case where $I^{(2)}_c$ does not depend on $N$. This is in accordance with the result of Ref.~\cite{zou2023channeling} which considers the Renyi negativity and the $g$ function. However, it is important to note that only Von-Neumann coherent information indicates the code properties. As we already stress, the $Z$ dephasing here offers an example which is marginal for $n\geq 2$ but irrelevant for $n\rightarrow 1$.

\end{document}